\documentclass[twocolumn,superscriptaddress,amsmath,amssymb,showpacs,prl]{revtex4-2}

\usepackage{graphicx}
\usepackage{color}
\usepackage{makecell}
\usepackage{footnote}
\usepackage[colorlinks=true, allcolors=blue]{hyperref}
\renewcommand{\phi}{\varphi}

\begin{document}

\title{Ferroelasticity, shear modulus softening, and the tetragonal\(\leftrightarrow \)cubic transition in davemaoite}

\author{Tianqi Wan}
	\affiliation{Department of Applied Physics and Applied Mathematics, Columbia University, New York, NY 10027, USA}

\author{Chenxing Luo}
	\affiliation{Department of Applied Physics and Applied Mathematics, Columbia University, New York, NY 10027, USA}
	\affiliation{Geosciences Department, Princeton University, Princeton, NJ, 08544, USA}

\author{Zhen Zhang}
	\affiliation{Department of Applied Physics and Applied Mathematics, Columbia University, New York, NY 10027, USA}
	\affiliation{Department of Physics, Iowa State University, Ames, 50011, USA}

\author{Yang Sun}
	\affiliation{Department of Physics, Iowa State University, Ames, 50011, USA}
	\affiliation{Department of Physics, Xiamen University, Xiamen 361005, China}

\author{Renata M. Wentzcovitch}
	\email{rmw2150@columbia.edu}
	\affiliation{Department of Applied Physics and Applied Mathematics, Columbia University, New York, NY 10027, USA}
	\affiliation{Department of Earth and Environmental Sciences, Columbia University, New York, NY 10027, USA}
	\affiliation{Lamont–Doherty Earth Observatory, Columbia University, Palisades, NY 10964, USA}

\date{\today}

\begin{abstract}

Davemaoite (Dm), the cubic phase of CaSiO$_3$-perovskite (CaPv), is a major component of the Earth’s lower mantle. Understanding its elastic behavior, including its dissolution in bridgmanite (MgSiO$_3$-perovskite), is crucial for interpreting lower mantle seismology. Using machine-learning interatomic potentials and molecular dynamics, we investigate CaPv’s elastic properties across the tetragonal\(\leftrightarrow \)cubic transition. Our equations of state align well with experimental data at 300 K and 2,000 K, demonstrating the predictive accuracy of our trained potential. We simulate the ferroelastic hysteresis loop in tetragonal CaPv, which has yet to be investigated experimentally. We also identify a significant temperature-induced shear modulus softening near the phase transition, characteristic of ferroelastic\(\leftrightarrow \)paraelastic transitions. Unlike previous elasticity studies, our softening region does not extend to slab geotherm conditions. We suggest that ab initio-quality computations provide a robust benchmark for shear elastic softening associated with ferroelasticity, a challenging property to measure in these materials.  

\end{abstract}

\maketitle

\section{I. Introduction}
The lower mantle, extending from the seismic discontinuity at $\sim$660 km depth to the core–mantle boundary (CMB) at $\sim$2,890 km, has been shown by seismic tomography models to contain large-scale velocity heterogeneities whose origins are still debated. 
Davemaoite (Dm), i.e., cubic CaSiO$_3$-perovskite (CaPv), comprises 7–10 vol\% in pyrolite\cite{Irifune1994,Saikia2008} and up to 30 vol\% in subducted mid-ocean ridge basalt (MORB) at depths 
below 560 km \cite{Hirose2002,Irifune1993,Kesson1994}. As the third most abundant phase in the lower mantle, its presence significantly affects this region’s properties\cite{Irifune1994} and those of subducted basaltic crust \cite{Hirose2002}. Experimental \cite{Irifune2000} and theoretical \cite{Vitos2006,Jung2011,Muir2021} studies suggest that the solubility 
of CaPv in bridgmanite (MgSiO$_3$-perovskite), the most abundant mantle phase, increases with temperature and is completely dissolved at $\sim$1,800 km depth along the geotherm \cite{Ko2022}. Furthermore, CaPv is an important Ti-bearing phase in subducted slabs, forming a Ca(Si,Ti)O$_3$ solid solution \cite{Chao2024,Thomson2016} with altered properties. It has 
been suggested that the key tetragonal\(\leftrightarrow \)cubic transition in Ti-bearing CaPv occurs at mid-mantle depths and may account for the seismological 
properties of large-low-shear-velocity provinces (LLSVPs) \cite{Thomson2019a}. Despite the potential influence of the Ca(Si,Ti)O$_3$ and (Mg,Ca)SiO$_3$-perovskite solid solutions on deep-mantle properties, the elastic behavior of pure CaPv under the mantle conditions 
remains highly relevant but debated and has yet to be fully addressed. Therefore, it is crucial to investigate its elastic properties first, particularly across the tetragonal\(\leftrightarrow \)cubic transition.

CaPv undergoes a tetragonal ($I4/mcm$) to 
cubic ($Pm\overline{3}m$) phase transition with increasing temperature above $\sim$12 GPa. Experimental \cite{Komabayashi2007,Kurashina2004} and theoretical studies \cite{Sagatova2021a,Li2006b,Caracas2005} of this transition 
have previously offered dissimilar phase boundaries. However, a recent large-scale computational study examining this phase boundary from a thermodynamic viewpoint \cite{Wu2024a} confirms detailed measurements, pinning this transition at sub-mantle-geotherm conditions. However, it remains unclear how this phase change may affect the elastic 
properties and acoustic velocities in CaPv at mantle conditions. The tetragonal\(\leftrightarrow \)cubic symmetry change in a perovskite system has long been predicted to exhibit ferroelastic behavior in the low-symmetry phase \cite{Carpenter1998b}, characterized by non-Hookean deformation with a hysteresis loop, similar to other ferroic properties such as ferroelectricity and ferromagnetism. As CaPv belongs to the same perovskite family, one may anticipate that tetragonal CaPv also exhibits ferroelastic behavior. However, no 
complete ferroelastic hysteresis loop in tetragonal CaPv has been observed, unlike the one first identified in 1976 in the prototypical ferroelastic material Pb$_3$(PO$_4$)$_2$ \cite{Salje1976}. Furthermore, strong velocity anomalies, characterized by a sharp shear velocity dip near the ferroelastic\(\leftrightarrow \)paraelastic 
transition, have been confirmed in several geophysically important species, including SiO$_2$, BiVO$_4$, and KMnF$_3$ \cite{Carpenter1998b,Zhang2021a}. Unfortunately, the challenges posed by CaPv’s spontaneous amorphization during decompression \cite{Liu1975} have limited similar studies in 
this material. The struggle to capture such hysteresis loops and elastic anomalies in ultrasound measurements emphasizes the importance of accurate computational studies to understand the seismological properties of CaPv and similar ferroelastic systems. Here, 
we apply deep learning techniques to generate interatomic potentials that accurately model CaPv’s compression curve and elastic properties at several temperatures. 
Exploring the hysteresis curve provides direct evidence of ferroelastic behavior in tetragonal CaPv. This method allows us to fully address the predicted yet previously unobserved shear modulus anomalies around the phase transition.

\section{II. Methods}
\subsection{2.1 AIMD simulations}
To create benchmark datasets for CaPv, Born-Oppenheimer molecular dynamics (BOMD) simulations were conducted using the Vienna Ab initio Simulation Package (VASP) \cite{chenxing30}. 
The effectiveness of the strongly constrained and appropriately normed (SCAN) meta-generalized gradient approximation (GGA) functional \cite{Sun2015} was carefully investigated. 
The simulations employed a $2 \times 2 \times 2$ supercell containing 40 atoms, alongside a $2 \times 2 \times 2$ \textbf{k}-point grid and a kinetic energy cutoff of 550 eV, which was shown to be sufficient 
to achieve convergence in calculating anharmonic phonon dispersions \cite{Zhang2021c}. One independent DP model, trained on SCAN datasets, was developed and is referred to as DP-
SCAN. The model training utilized the smooth version descriptor $\texttt{se\_e2\_a}$ proposed by Zhang et al \cite{Zhang2018b,DPMD} (see Text S1 in Supplementary Information).

\subsection{2.2 DPMD simulations}
We conducted deep potential molecular dynamics (DPMD) simulations using the LAMMPS code \cite{Thompson2022}. An $NPT$ simulation was first performed to equilibrate the cell shape under 
specific temperature and pressure conditionss. This simulation involved 10,240 atoms, ran for 0.1 ns, and used a time step of 0.2 fs. The Nosé-Hoover thermo-baro-stat that 
incorporates modular invariance \cite{Hoover1996,Wentzcovitch1991a} were applied for temperature and pressure control. After the system reached equilibrium, subsequent simulations started from the 
equilibrated unit cell shape. To determine the elastic tensor through the stress-
coordinates and velocities at target temperatures. These were followed by the 4 ns-long $NVE$ simulations with a timestep of 0.5 fs to compute the adiabatic $\textbf{c}_{ij}^s$ tensor components under defined conditions.
\subsection{2.3 The stress-fluctuation formalism for thermoelasticity}
The thermodynamic elastic tensor \( A_{ijkl} (T, V) \) (where \( i, j, k, l = 1, 2, 3 \)) represents the second derivative of the free energy \( F \) density with respect to strain under equilibrium conditions, while the volume \( V \) is fixed in molecular dynamics (MD) simulations. It is expressed as \cite{Ray1984,Zhen2012,Clavier2023}:
\begin{align}
&A_{ijkl}(T, V) = \frac{1}{V} \frac{\partial^2 F}{\partial \varepsilon_{ij} \partial \varepsilon_{kl}} \notag \\ 
&= \langle A_{ijkl}^B \rangle - \frac{k_B T}{V} \left[ \langle \sigma_{ij} \sigma_{kl} \rangle - \langle \sigma_{ij} \rangle \langle \sigma_{kl} \rangle \right], \notag
\end{align}
, where \( \langle \cdot \cdot \cdot \rangle \) denotes the ensemble average, \( \varepsilon_{ij} \) is the strain tensor, \( N \) is the number of atoms, \( T \) is the temperature, and \( \delta_{ij} \) is the Kronecker delta function. The stress tensor \( \sigma_{ij} \) and Born matrix tensor \( A_{ijkl}^B \) are derived from the first and second derivatives of strain energy density \( U/V \) with respect to \( \varepsilon_{ij} \), respectively:
\begin{equation}
\sigma_{ij} = \frac{1}{V} \frac{\partial U}{\partial \varepsilon_{ij}}, \quad A_{ijkl} = \frac{1}{V} \frac{\partial^2 U}{\partial \varepsilon_{ij} \partial \varepsilon_{kl}}. \notag 
\end{equation}
During the equilibrated MD runs, both \( A_{ijkl}^B \) and \( \sigma_{ij} \) are periodically calculated. Since the convergence of \( A_{ijkl}^B \) is faster than that of \( \sigma_{ij} \), different 
sampling frequencies are used to optimize storage while maintaining computational accuracy (calculating \( A_{ijkl}^B \) every 200 MD steps and \( \sigma_{ij} \) every 20 MD steps). The Born matrix \( A_{ijkl}^B \) is computed as follows \cite{Clavier2023}:
\begin{equation}
A_{ijkl}^B = \frac{\partial \sigma_{ij}}{\partial \varepsilon_{kl}} = \sigma_{ij} \delta_{kl} + \sigma_{ik} \delta_{jl}. \notag
\end{equation}
For an average stress \( \sigma_{ij} \) calculated during the simulation, the effective elastic tensor \( C_{ijkl} \) is given by the following expression \cite{Barron1965,Luo2022}:
\begin{align}
&C_{ijkl} = \frac{1}{V} \frac{\partial^2 F}{\partial \varepsilon_{ij} \partial \varepsilon_{kl}} - \sigma_{ij} \delta_{kl}  \notag \\
&+\frac{1}{2} \left( \sigma_{ik} \delta_{jl} + \sigma_{kj} \delta_{il} + \sigma_{il} \delta_{jk} + \sigma_{lj} \delta_{ik} \right). \notag
\end{align}
Under hydrostatic stress condition (i.e., \( \sigma_{ij} = P \delta_{ij} \), where \( P = -\frac{1}{3} \text{Tr}(\sigma_{ij}) \)), this equation can be further simplified to \cite{Barron1965}:
\begin{equation}
C_{ijkl} = \frac{1}{V} \frac{\partial^2 F}{\partial \varepsilon_{ij} \partial \varepsilon_{kl}} +P\left(\delta_{ij}\delta_{kl}-\delta_{il}\delta_{kj}-\delta_{ik}\delta_{jl} \right). \notag
\end{equation}
Here, \( A_{ijkl} \) and \( C_{ijkl} \) are equal only when the system is free from external pressure or stress. The adiabatic elastic tensors \( A_{ijkl}^S \) and \( C_{ijkl}^S \) can be obtained from NVE simulations.

\section{III. Results}
\subsection{3.1 Ferroelastic behavior and hysteresis loop}
Molecular dynamics simulations using a deep learning potential trained with the strongly constrained and appropriately normed meta–generalized-gradient approximation 
functional (DFT-SCAN) \cite{Sun2015} can accurately reproduce the high-pressure elastic behavior of minerals, even across complex phase transitions \cite{Luo2024}. Comparison of our predicted 300 K DP-SCAN compression curve with experimental measurements \cite{Chen2018,Greaux2019a,Mao1989,Shim2002,Thomson2019a,Wang1996} is shown in 
Fig. 1A. We reproduce most of the measurements with unprecedented accuracy, with discrepancies smaller than 1\%, except for data from Ricolleau et al. \cite{Ricolleau2010a}. 
\begin{figure}
    \centering
    \includegraphics[height=17pc]{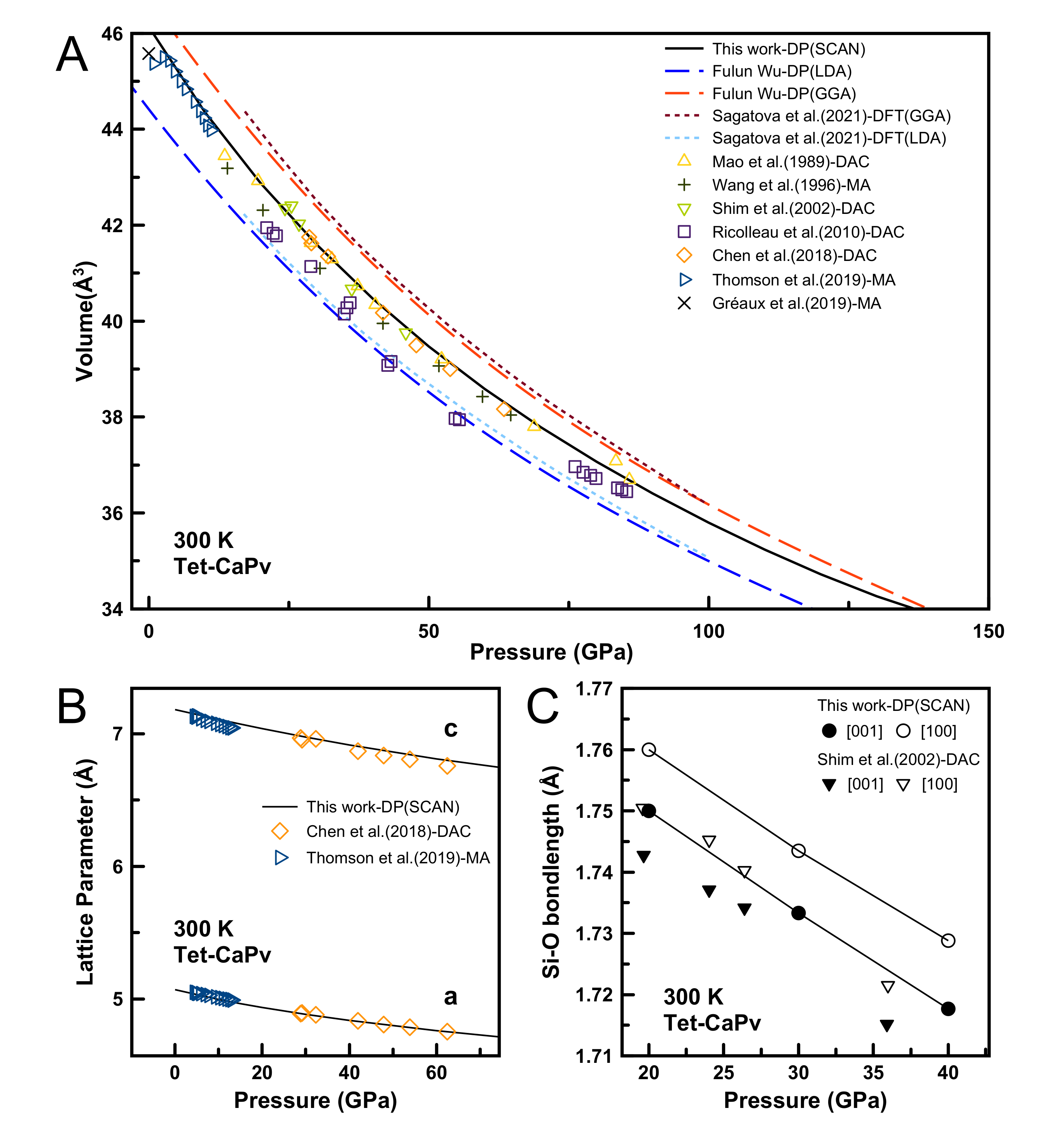}
    \caption{ (A) Compression curves for tetragonal CaPv at 300 K. Solid curves are obtained using DP-SCAN, dashed lines are previous DFT results \cite{Sagatova2021a,Wu2024a}, and experimental data are taken from Refs.\cite{Chen2018,Greaux2019a,Mao1989,Ricolleau2010a,Shim2002,Thomson2019a,Wang1996}. (B) Lattice parameters of CaPv as a function of pressure at 300K. (C) Si-O bond lengths in CaPv as a function of pressure at 300K. Solid symbols denote Si-O bond along \textbf{c}-axis, while open symbols denote Si-O bond along \textbf{a}-axes.}\label{fig1}
\end{figure}
This potential also reliably reproduces axial compressibility (Fig. 1B) and Si–O bond 
lengths (Fig. 1C). Notably, the Si–O bond length along the [001] direction remains slightly longer than along [100]. Compared to previous local density approximation (LDA) and generalized gradient approximation (GGA) based predictions, SCAN reproduces 
these structural properties more accurately. Except for Ricolleau et al. \cite{Ricolleau2010a} data, the compressive behavior of cubic CaPv measured by X-ray diffraction experiments at 2,000 K was also well-reproduced in our simulations up to 120 GPa (see Fig. S2 in Supplementary Information). This 
agreement indicates that DP-SCAN describes the behavior of both tetragonal and cubic CaPv well at relevant conditions, validating the DP-SCAN method to explore the CaPv system throughout a broad range of conditions.

\begin{figure*}
\centerline{\includegraphics[width=360pt]{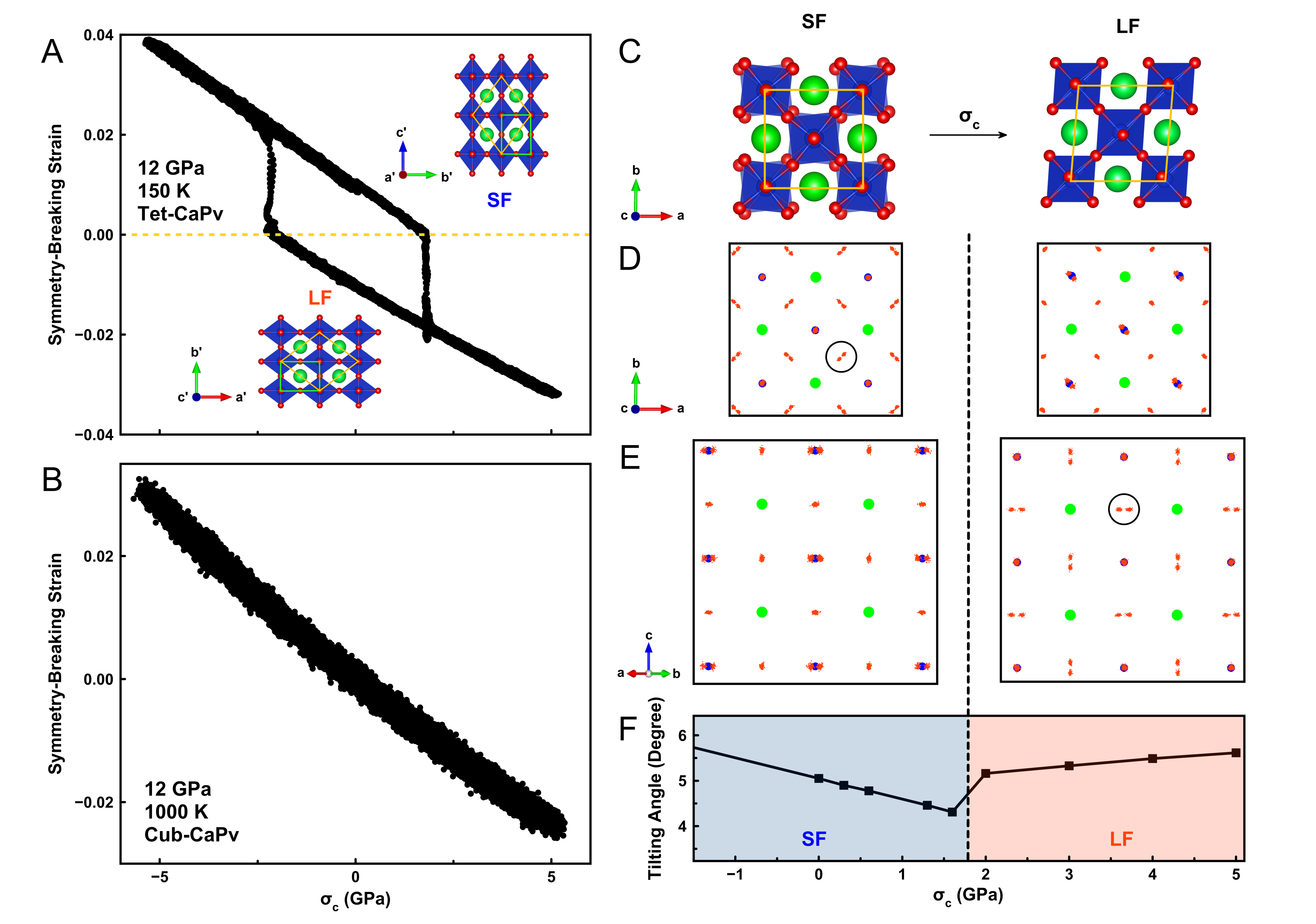}}
\caption{(A) Ferroelastic behavior with hysteresis loop in tetragonal CaPv at 12 GPa and 150 K. Green squares in the inset structures denote the corresponding cubic CaPv primitive cell ($1\times1\times1$, 5 atoms), while yellow lozenge represents the tetragonal CaPv primitive cell ($\sqrt{2} \times \sqrt{2} \times 2$, 20 atoms). (\textbf{a}', \textbf{b}', \textbf{c}') represent the lattice vectors of the cubic cell, while (\textbf{a}, \textbf{b}, \textbf{c}) denote the lattice vectors of the tetragonal cell. (B) Paraelastic behavior in cubic CaPv at 12 GPa and 1,000 K. Uniaxial stress was applied along \textbf{c}-axis from -5 GPa (tensile) to 5 GPa (compression) gradually (see Text S2 in Supplementary Information). (C) The crystal structure of tetragonal CaPv projected on the (001) plane of SF state and LF state. Blue, red, and green denote Si, O, and Ca ions; (D) The trajectories of O ions projected on the (001) plane; (E) The trajectories of O ions projected on the (110) plane; (F) SiO$_6$ octahedra tilting angles w.r.t. to the longest axes in the SF and LF states.}\label{fig2}
\end{figure*}

One of the key findings in this study is that tetragonal CaPv exhibits elastic hysteresis characteristic of ferroelastic systems under a stretch-compression cycle 
with uniaxial stress applied along the \textbf{c}-axis (Fig. 2A). Such a feature was not seen in prior simulations or experiments whereas is consistent with the ferroelastic behavior 
observed in other perovskites \cite{Carpenter1998b}. At 150 K and 12 GPa, compressive stress along the \textbf{c}-axis causes the symmetry-breaking strain $e_t=[(c-a)/a]$ to approach 0, indicating a transition towards a cubic-like structure. However, the cubic structure is unstable under these conditions, leading to 
a switch from a standing form (SF) to a lying-down form (LF) (Fig. 2A), corresponding to different orientations of the longer side of the tetragonal phase (see Text S2 in Supplementary Information). Further analysis of the trajectories of oxygen ions reveals that SiO$_6$ octahedra 
rotate around the \textbf{c}-axis in the SF state and around the [110] direction in the LF state (Fig. 2D and E). In the SF state, the tilting angle decreases with increasing compressive stress. In contrast, in the LF state, the tilting angle increases with 
compressive stress along the original \textbf{c}-axis (Fig. 2F) since the compressive stress axis and the longer length axis are now perpendicular. The mechanical switching between these two states under external compressive stress results from a shear strain within 
the \textbf{ab}-plane of the tetragonal cell, as illustrated in Fig. 2C. Moreover, as temperature approaches the transition temperature ($T_{tr}$), \textbf{c} and \textbf{a} approach similar values, and smaller stress perturbations can produce this mechanical switching, potentially 
explaining the pronounced dip in the shear modulus near $T_{tr}$. Microscopically, the system’s stress vs. strain relation still follows Hooke’s law before and after the switching (Fig. 2A). Also, the ferroelastic hysteresis loop’s amplitude depends significantly on temperature. At 150 K and 12 GPa, the induced stress required to change the longer axis 
orientation is $\sim$1.7 GPa. As the temperature approaches $T_{tr}$, the hysteresis loop shrinks and eventually disappears (see Fig. S3 in Supplementary Information). This is the archetypical behavior of a ferroelastic material undergoing a transition from a low-symmetry ferroelastic phase 
to a high-symmetry paraelastic one \cite{Salje2012}, $I4/mcm$ to $Pm\overline{3}m$ in this case. As shown in Fig. 2B, the ferroelastic hysteresis loop disappears entirely in cubic CaPv, the paraelastic counterpart. Ferroelastic\(\leftrightarrow \)paraelastic transitions occur naturally in oxides and 
silicates within Earth’s crust and mantle and have been reported to cause seismic velocity anomalies \cite{Kaneshima2016,Salje1992}. The effects of ferroelastic transitions on seismic parameters \cite{Zhang2021a} are expected to differ significantly from those caused by structural 
transitions and temperature-compositional variations more commonly observed in the mantle \cite{Zhuang2024a}. To better understand the nature of these transitions, we further investigate the complete set of elastic moduli across the transition and a broad range of pressures and temperatures around it.

\subsection{3.2 Softening region near the phase transition}
CaPv undergoes a ferroelastic\(\leftrightarrow \)paraelastic transition at $T=T_{tr}$ similar to other perovskites such as BaTiO$_3$ and SrTiO$_3$ \cite{Carpenter1998b}. Individual components of the elastic tensors (Fig. 3) 
display strong anomalies near $T_{tr}$. The three sets of elastic coefficients for cubic CaPv, i.e., diagonal ($\textbf{c}_{11}$ = $\textbf{c}_{22}$ = $\textbf{c}_{33}$), shear ($\textbf{c}_{44}$ = $\textbf{c}_{55}$ = $\textbf{c}_{66}$), and off-diagonal ($\textbf{c}_{12}$ = $\textbf{c}_{13}$ = $\textbf{c}_{23}$), gradually diverge at $T_{tr}$, indicating that
the cubic lattice experiences increasingly anisotropic contraction and shear strains as temperature decreases. In tetragonal CaPv, the shear elastic coefficient 
$\textbf{c}_{66}$ associated with 
shear strain \(\boldsymbol{\varepsilon}_{xy}\), has a 
significantly larger value than the $\textbf{c}_{44}$ or $\textbf{c}_{55}$ coefficients.
The smaller $\textbf{c}_{44}$ and $\textbf{c}_{55}$ values can be attributed to the 
rotation of SiO$_6$ octahedra along the longer axis. This rotation would result from the lattice distortion transition where the cubic \textbf{a}-axis of CaPv splits into the 
tetragonal \textbf{a}- and \textbf{c}-axis. The three principal longitudinal coefficients follow the trend $\textbf{c}_{11}$ $>$ $\textbf{c}_{33}$, which indicates anisotropic lattice distortions. The gradual symmetry reduction across the 
transition with a symmetry-breaking strain is usually accompanied by a dramatic anomaly of the shear modulus near $T_{tr}$.

\begin{figure}
    \centering
    \includegraphics[height=20pc]{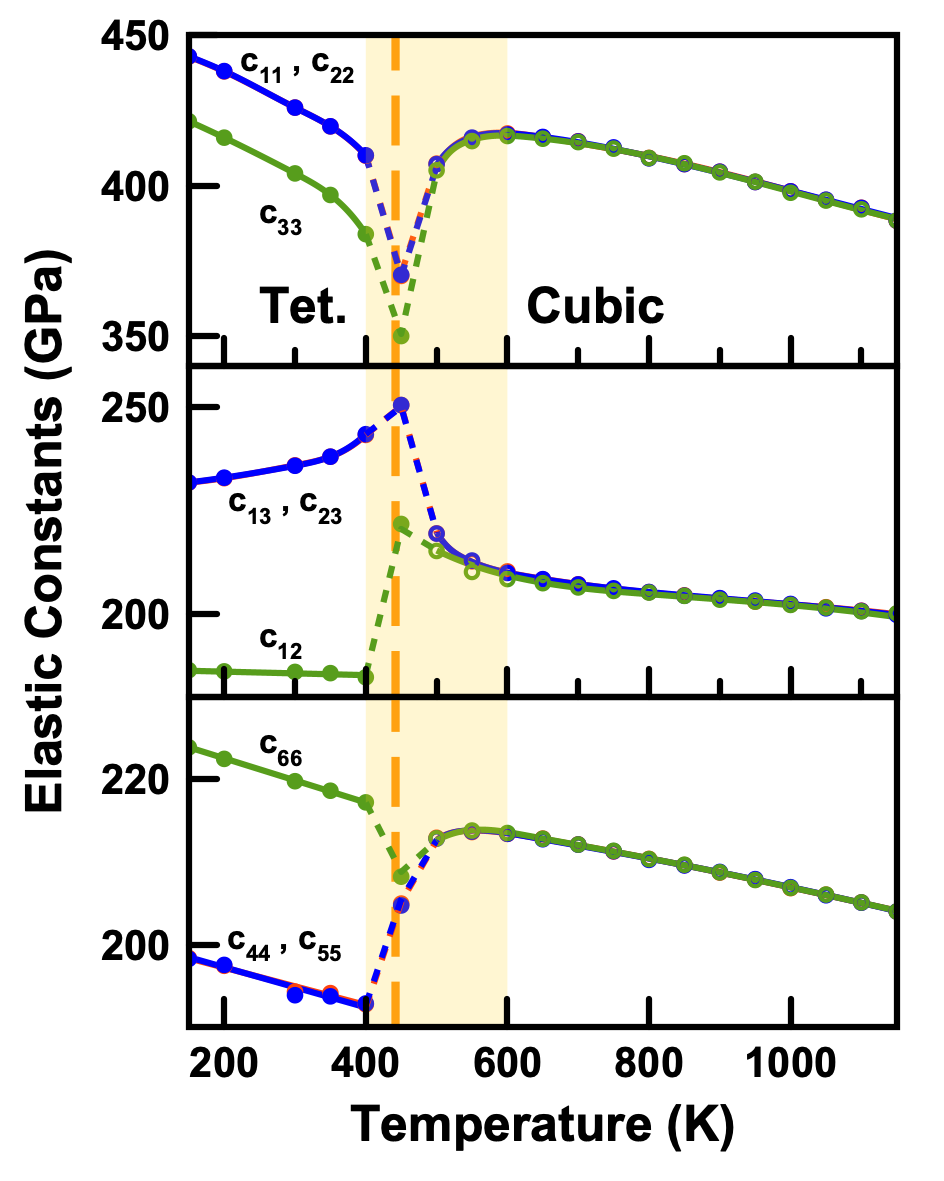}
    \caption{Elastic constants vs. temperature at 12 GPa obtained using the DP-SCAN potentials and methods listed under Materials and Methods. Orange dashed lines indicate the phase transition measured by Thomson et al. \cite{Thomson2019a}. The dashed lines connecting the elastic coefficients of both phases to the minimum values are for guidance to the eye. The yellow-shaded area shows the “G-softening” region. }\label{fig3}
\end{figure}

\begin{figure}
    \centering
    \hspace{-0.5cm}
    \includegraphics[height=23pc]{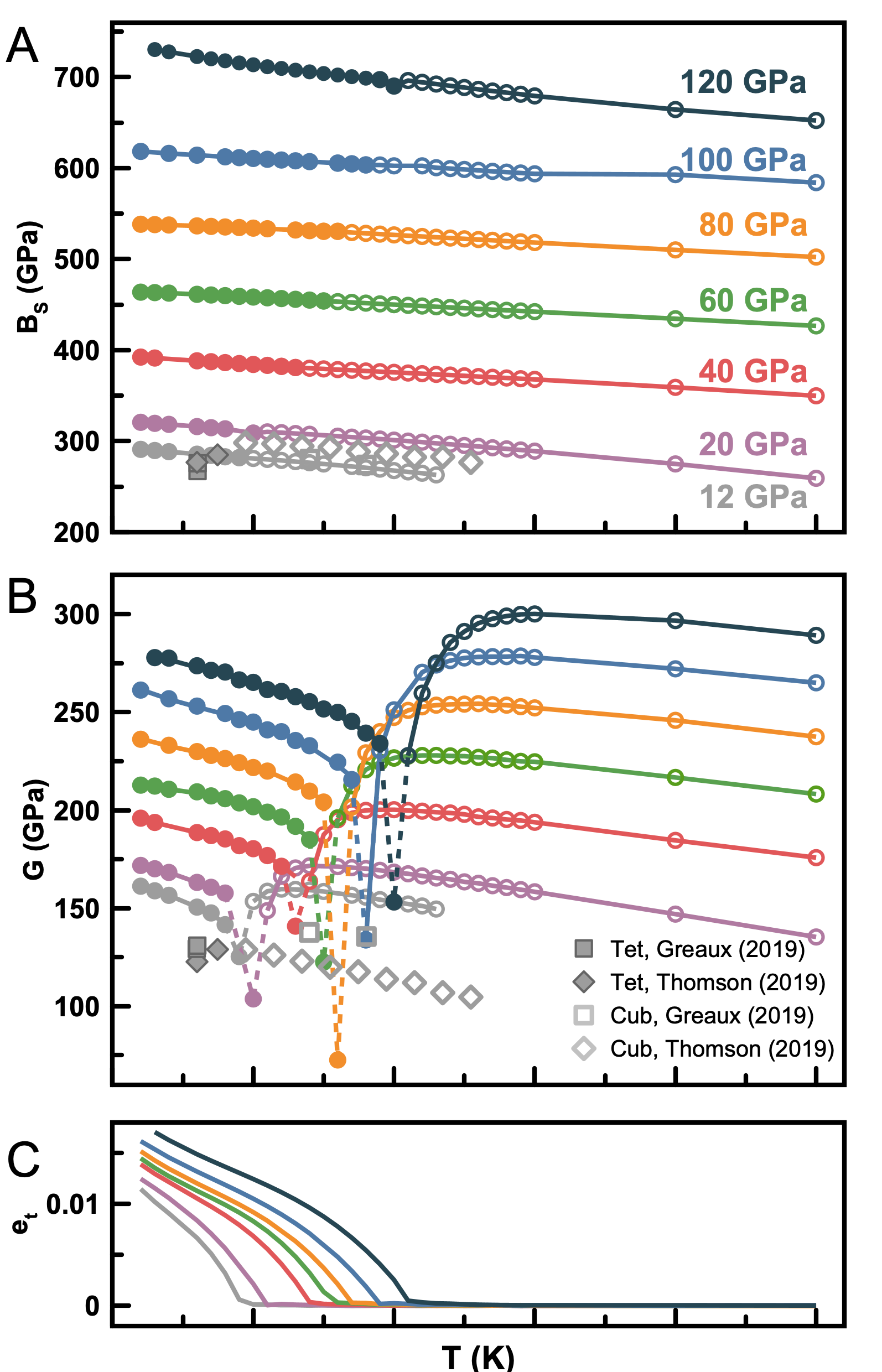}
    \caption{(A) Bulk modulus-$B_S$, (B) shear modulus-$G$, and (C) symmetry-breaking strain of CaPv-$e_t$ as a function of temperature. Measured elastic moduli from previous studies are plotted as grey diamonds \cite{Thomson2019a} (about 12 GPa) and grey squares \cite{Greaux2019a} (12 ± 1 GPa). }\label{fig4}
\end{figure}

\begin{figure*}
\centerline{\includegraphics[width=360pt]{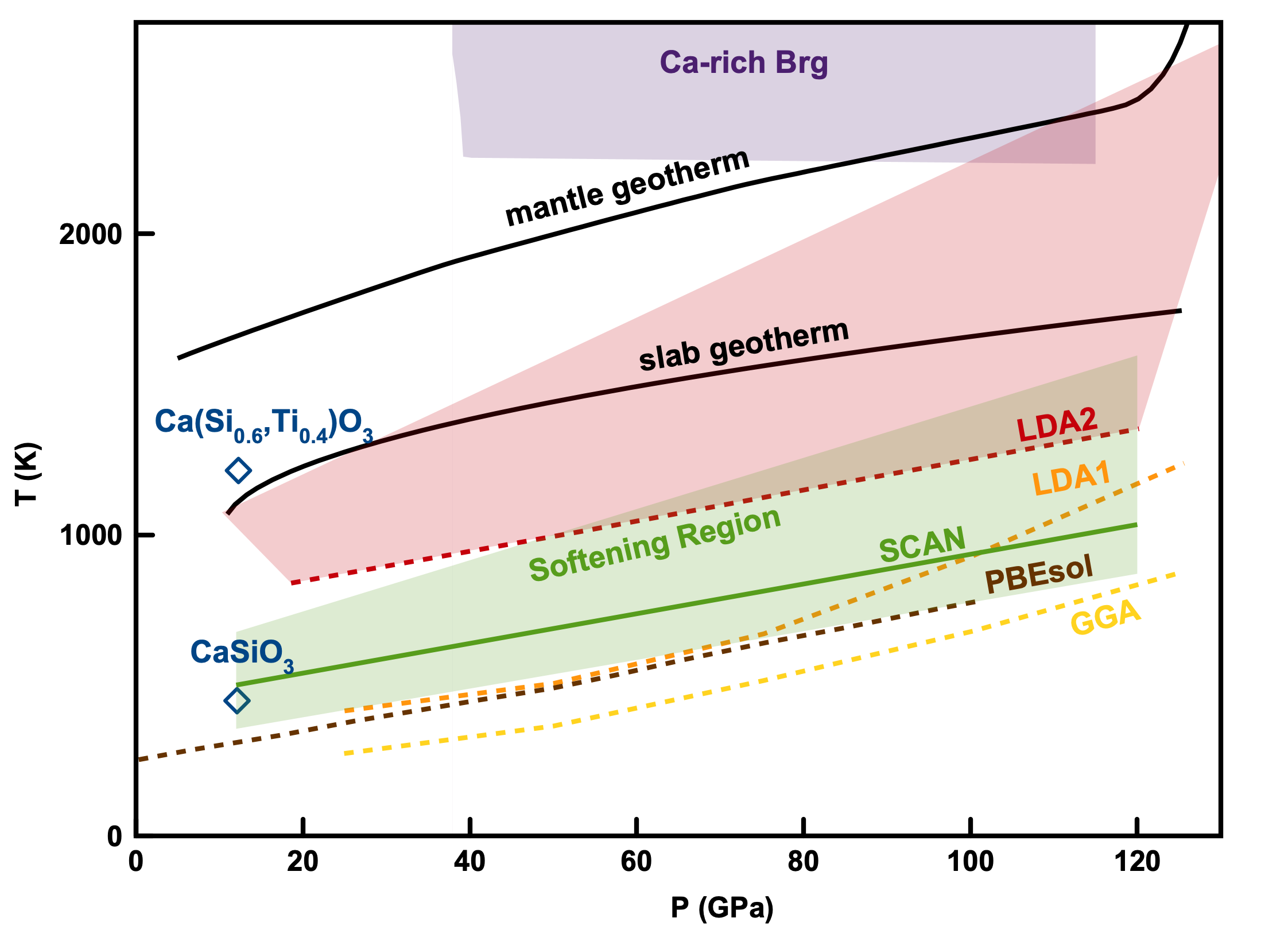}}
\caption{Phase diagram of CaPv. Normal mantle \cite{Brown1981} and cold slab \cite{Eberle2002} geotherms are plotted also. The green line is the current DP-SCAN results corresponding to the mid-point between the minimum values of $G$ in tetragonal and cubic CaPv in Fig. 4B. The green-shaded area shows the “$G$-softening” region in this study, while the red-shaded area shows the results from Zhang et al. \cite{Zhang2025}. Previous boundaries calculated using DP-GGA (dashed-yellow) and DP-LDA (LDA1) (dashed-orange) \cite{Wu2024a}, PBEsol (dashed-brown) \cite{Shin2024}, and machine-learning force fields-LDA (LDA2) (dashed-red) \cite{Zhang2025} are also shown. Dark blue diamonds represent measurements with different titanium concentrations \cite{Thomson2019a}. The purple-shaded region shows the observed stability field of Ca-rich bridgmanite \cite{Ko2022}.}\label{fig5}
\end{figure*}

To further evaluate CaPv’s elastic properties across the ferroelastic\(\leftrightarrow \)paraelastic transition, we analyzed the behavior of the adiabatic bulk modulus ($B_S$) and shear 
modulus ($G$) computed using Voigt-Reuss-Hill (VRH) averages (Fig. 4A and B). $B_S$ aligns closely with previous measurements \cite{Thomson2019a,Greaux2019a} at 12 GPa and retains a nearly linear dependence in tetragonal and cubic CaPv. Additionally, the modest reduction in $B_S$ near $T_{tr}$ is consistent with prior observations (see Fig. S4 in Supplementary Information). The less accentuated reduction observed in our calculations may result from domain structures present in experimental measurements but absent in simulations with 10,240 atoms. A 
recent similar simulation does not show this small anomaly in $B_S$ near $T_{tr}$, likely because it examined the elastic anomaly only in the high-pressure cubic phase. Notably, the transition does not cause significant changes in density \cite{Stixrude2007,Tsuchiya2011} (see Fig. 
S5 in Supplementary Information). In contrast to $B_S$, a very strong reduction in $G$ is revealed by calculations compared to measurements at 12 GPa. The characteristic dip in $G$ occurs in the 
ferroelastic phase when the symmetry-breaking strain $e_t$ approaches zero around the ferroelastic\(\leftrightarrow \)paraelastic transition. An even stronger $G$ softening occurs in the parealastic phase near $T_{tr}$. 
Our predicted phase boundary is lower than the latest computational estimate by Zhang et al. \cite{Zhang2025}. We used different but equivalent approaches to outline the phase boundary—tracking elastic anomaly vs. tracking lattice parameters in theirs (see Figs. 4B and 4C). Besides, our elastic properties agree well with theirs (see Fig. S6 in Supplementary Information). Therefore, the difference in our phase boundaries should be primarily due to using different DFT functionals. Zhang et al. \cite{Zhang2025} used LDA with an approximate 4.3 GPa pressure shift, whereas we employed SCAN, which produces the most accurate EOS to date (see Figs. 1 and S2). In contrast, phase boundaries obtained by comparing Gibbs free energies and computed using LDA, GGA \cite{Wu2024a}, and PBEsol \cite{Shin2024} yield lower transition temperatures, which is not surprising for ferroelastic systems \cite{Carpenter1998b}. These observations suggest that the SCAN-derived phase boundary obtained by tracking elastic anomalies or lattice parameters can differ slightly from the true thermodynamic SCAN boundary, which is expected to be the most reliable but has yet to be computed.
Notably, $G$ exhibits a nearly linear temperature dependence well above and below $T_{tr}$,but displays nonlinear behavior over a broad 
temperature range around  $T_{tr}$ (yellow shaded area in Fig. 5). In this nonlinear region, the ``softening region'' hereafter, $G$ exhibits different degrees of softening depending on 
pressure. With increasing pressure, the softening temperature range extends 300–700 K around $T_{tr}$. Therefore, ferroelastic transitions can anomalously impact seismic parameters 
over a broad temperature range far beyond that of a sharp thermodynamic phase boundary. Although the $G$ softening temperature range broadens with increasing pressure, it 
remains below the cold slab geotherm (Fig. 5) where CaPv is expected to be abundant. This suggests that the significant softening associated with the tetragonal\(\leftrightarrow \)cubic 
transition in pure CaPv alone is unlikely to cause seismic wave speed anomalies in the lower mantle. This result contrasts with previous elasticity results \cite{Zhang2025} that extend 
the softening region to the slab geotherm conditions. This discrepancy is due to their overestimation of the tetragonal\(\leftrightarrow \)cubic transition temperature (see Fig. 5).

\begin{figure}
    \centering
    \hspace{-0.5cm}
    \includegraphics[height=21pc]{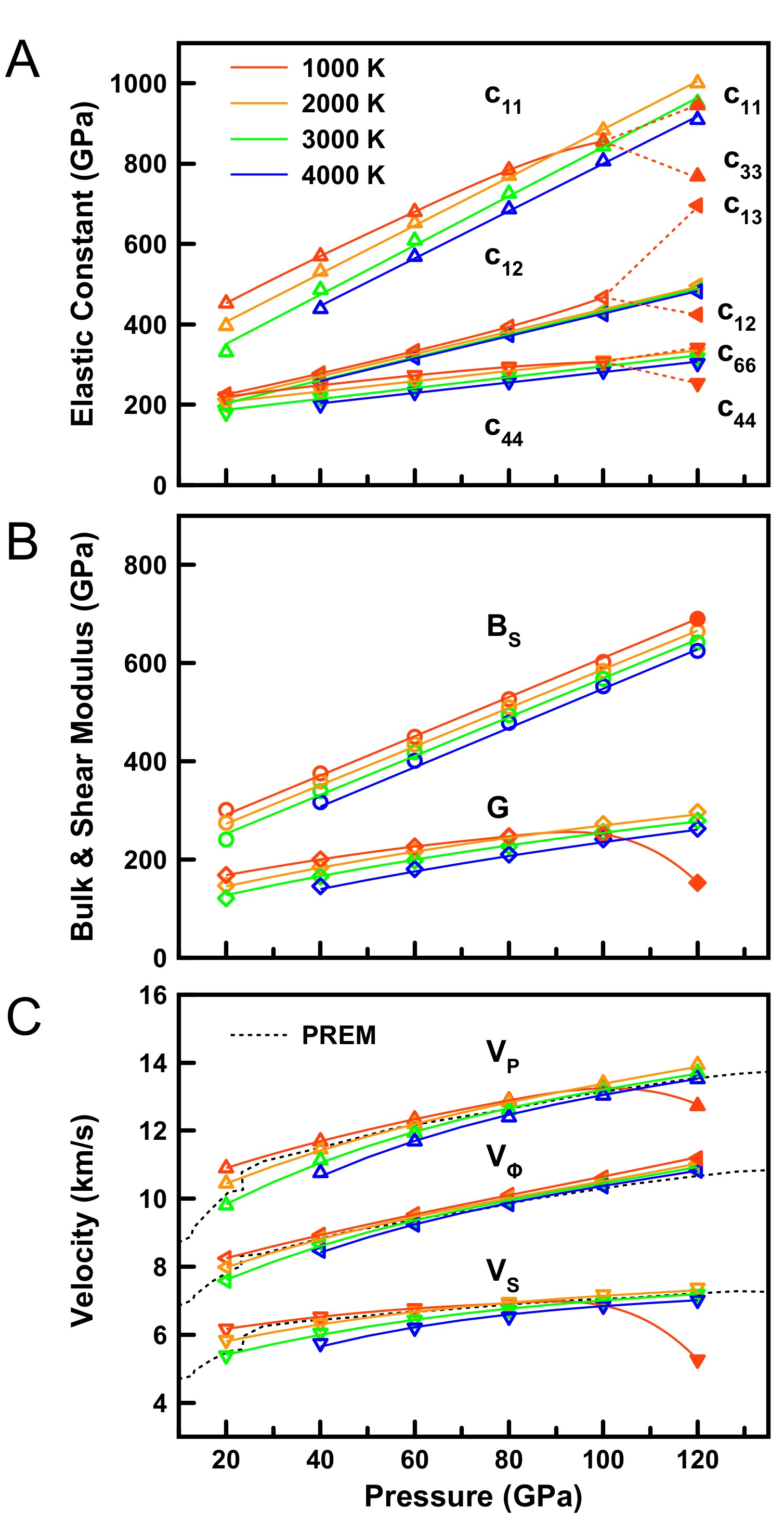}
    \caption{(A) Adiabatic elastic constants of CaPv vs. pressure at various temperatures calculated using stress-fluctuation formalism. Filled symbols denote the tetragonal CaPv, while open symbols denote the cubic CaPv. Fitting curves for each temperature are shown in lines.  (B) Averaged bulk and shear moduli and (C) Longitudinal ($v_P$), bulk sound ($v_{\phi}$), and shear ($v_S$) wave velocities of CaPv at same temperatures. Black dashed lines indicate the PREM values \cite{Dziewonski1981}.}\label{fig6}
\end{figure}

Fig. 6 shows cubic CaPv’s elastic and seismic properties across various temperatures ($T$) and pressures ($P$). Fig. 6B shows a typical behavior of $B_S$ vs. $P$ and $T$, similar to 
bridgmanite’s bulk modulus (e.g., \cite{Wan2024}). In contrast, $G$ displays a strong no-linear $P$-dependence at lower $T$s. For example, consistent with previous measurements (\cite{Greaux2019a}, the $G$ 
and $V_S$ reduction at 1,000 K are less dramatic vs. $P$ than vs. $T$. At conditions where the elastic coefficients deviate significantly from nearly linear $T$-dependence (Fig. 6A), $G$ and wave velocities (Fig. 6B and Fig. 6C) do as well. In summary, close to the ferroelastic\(\leftrightarrow \)paraelasatic transition (yellow shaded area in Fig. 5), the cross 
derivative,  \(\frac{\partial^2 D}{\partial T \partial P}\) (with $D$ being an elastic coefficient, $G$, or wave velocity), displays clear $T$- and $P$-dependences.

\section{IV. Disscussion}

Seismic velocity heterogeneities and anomalies, often attributed to temperature or compositional variations, could also be influenced by the presence of ferroelastic 
domains near $T_{tr}$. Specifically, the presence of a significant amount of Ti in MORB produces a Ca(Si,Ti)O$_3$ solid solution that stabilizes the tetragonal phase to higher 
temperatures, potentially extending the effects of the tetragonal\(\leftrightarrow \)cubic transition to mantle conditions \cite{Thomson2016,Thomson2019a}. In particular, it has been suggested that the seismological 
properties of large-low-shear-velocity-provinces, LLSVPs, are consistent with a moderate enrichment of MORB containing large amounts of Ca(Si,Ti)O$_3$  in these regions 
\cite{McNamara2005,Brandenburg2007}. Besides, the recent prediction \cite{Muir2021} and observation \cite{Ko2022} of the dissolution of CaPv in bridgmanite raise questions about how the seismological properties of the most 
abundant mantle phase, bridgmanite, might be affected by ferroelasticity in the CaPv endmember.

Before such questions can be addressed, a clearer understanding of the seismological properties of pure CaPv must emerge, which might be difficult for many reasons, 
including challenging measurements at appropriate conditions. Ferroelastic phases 
display a well-known sound velocity dependency on frequency (dispersion) near the ferroelasti\(\leftrightarrow \)cparaelastic transition \cite{Kohutych2010}, with this dependency arising from strain-induced ferroelastic domain wall motion. This might have been the source of subtle 
differences between ultrasound measurements in CaPv at 30$\sim$60 MHz \cite{Thomson2019a,Greaux2019a}. Domain wall motion greatly enhances elastic softening w.r.t. to that caused by the phase transition 
alone \cite{Carpenter2007}. One can expect that seismic waves with much lower frequencies (0.001 to 1 MHz) may experience even more reduced velocities near $T_{tr}$. Additional anelastic effects 
beyond ferroelasticity, e.g., grain boundary and creep, which are extremely challenging to compute, are also expected to affect wave velocities with such frequencies. In 
contrast, \textit{ab initio} quality calculations such as those performed here provide a controlled and reliable approach to estimating elastic softening related to 
ferroelasticity alone. They provide a baseline by isolating purely elastic contributions free from frequency-dependent anelastic effects throughout variable 
thermodynamic conditions. Brillouin scattering measurements might be able to test these predictions. While experimental measurements and computations have yet to advance to 
accurately capture or predict fully dispersive acoustic wave speeds under geophysically relevant conditions, their interplay provides a deeper understanding of ferroelastic 
material properties. This interaction highlights the importance of integrating methodologies to address the subtle complexities of ferroelasticity and its potential 
contributions to the overall anelastic behavior of minerals under the extreme conditions of Earth's interior.

\section{Acknowledgments}
This work was supported by DOE Award DE-SC0019759. R.M.W also acknowledges support from the Gordon \& Betty Moore Foundation (Moore 12801). Y.S. acknowledges support from the 
NSFC (Grants Nos. T2422016 and 42374108). Calculations were performed on the Extreme Science and Engineering Discovery Environment (XSEDE) (Towns et al., 2014) supported by 
the NSF Grant 1548562 and Advanced Cyberinfrastructure Coordination Ecosystem: Services \& Support (ACCESS) program, which is supported by NSF Grants 2138259, 2138286, 
2138307, 2137603, and 2138296 through allocation TG-DMR180081. Specifically, it used the Bridges-2 system at the Pittsburgh Supercomputing Center (PSC), the Anvil system at 
Purdue University, the Expanse system at San Diego Supercomputing Center (SDSC), and the Delta system at National Center for Supercomputing Applications (NCSA).

\twocolumngrid

\bibliographystyle{apsrev4-1}


\pagebreak
\widetext
\begin{center}
\textbf{\large Supplementary Information for Ferroelasticity, shear modulus softening, and the tetragonal\(\leftrightarrow \)cubic transition in davemaoite}
\end{center}

\setcounter{equation}{0}
\setcounter{figure}{0}
\setcounter{table}{0}
\renewcommand{\thefigure}{S\arabic{figure}}
\renewcommand{\thetable}{S\arabic{table}}

\textbf{This  file includes:}\\

Supplementary text S1 to S2\\

Figures S1 to S7\\

Tables S1 \\

\section{\textbf{Supplementary Text S1: Development of Machine-Learning Potential}}
The Neural Network potential for Davemaoite (Dm), i.e., cubic CaSiO$_3$-perovskite (CaPv), was developed based on the Deep Potential Smooth Edition (\texttt{DEEPPOT-SE}) model \cite{Zhang2018b} implemented in \texttt{DEEPMD-KIT} v2.1 \cite{Wang2018,chenxing37}. Two-body embedding with coordinates of the neighboring atoms (\texttt{se\_e2\_a}) was used for the descriptor. The embedding network was designed with a shape of (25, 50, 100), while the fitting network had a shape of (240, 240, 240). The cutoff radius is 6 \AA, and the smoothing parameter of 0.5 \AA. The model was trained using the Adam optimizer \cite{Kingma2014} for $1 \times 10^6$ training steps, with the learning rate exponentially decaying from $1 \times 10^{-3}$ to $3.51 \times 10^{-8}$ during the training process. The loss function \( \mathcal{L}(p_e, p_f) \) is given by \cite{Wang2018}:
\begin{equation}
\mathcal{L}(p_e, p_f) = p_e |\Delta e|^2 + \frac{p_f}{3N}|\Delta f_i|^2,
\end{equation}
where \( p_e \) linearly decays from 1.00 to 0.02, while \( p_f \) linearly increases from \( 1 \times 10^0 \) to \( 1 \times 10^3 \) throughout the training process.

We employed the DP-GEN concurrent learning scheme to create the reference dataset and generate the potential \cite{Zhang2020}. Initially, we randomly extracted 200 labeled configurations from 50 MD runs spanning various $P$-$T$ ranges, covering 0-160 GPa and 300-4000 K to generate the initial potentials. We performed four DP-GEN iterations to explore the configuration space and ultimately achieve a potential that meets the desired accuracy for MD simulations. We trained four candidate DP potentials initialized with different random seeds in each iteration. The error estimator (model deviation) \( \epsilon_t \) is determined based on the force disagreement between the candidate DPs \cite{Zhang2019, Zhang2020}. The expression for \( \epsilon_t \) is:
\begin{equation}
\epsilon_t = \max_i \sqrt{\langle\left\lVert F_{\omega,i}(\mathcal{R}_t) - \langle F_{\omega,i}(\mathcal{R}_t) \rangle \right\rVert^2\rangle} ,
\end{equation}
where \( F_{\omega,i}(\mathcal{R}_t) \) represents the force on the \( i \)-th atom predicted by the \( \omega \)-th potential for the configuration \( \mathcal{R}_t \). For a particular configuration, if \( \epsilon_t \) satisfies \( \epsilon_{\text{min}} \leq \epsilon_t \leq \epsilon_{\text{max}} \), the corresponding configuration is collected, then labeled with DFT forces and total energy then added into the training data set; if \( \epsilon_t < \epsilon_{\text{min}} \), these configurations are considered covered by the current training data set; if \( \epsilon_t > \epsilon_{\text{max}} \), these configurations are considered failed discarded. After a few iterations, almost no new configurations are collected according to this standard (\( >99\% \) are accurate for a few iterations) and the DP-GEN process is then complete.

\begin{figure}
\centering
	\includegraphics[width=0.7\textwidth]{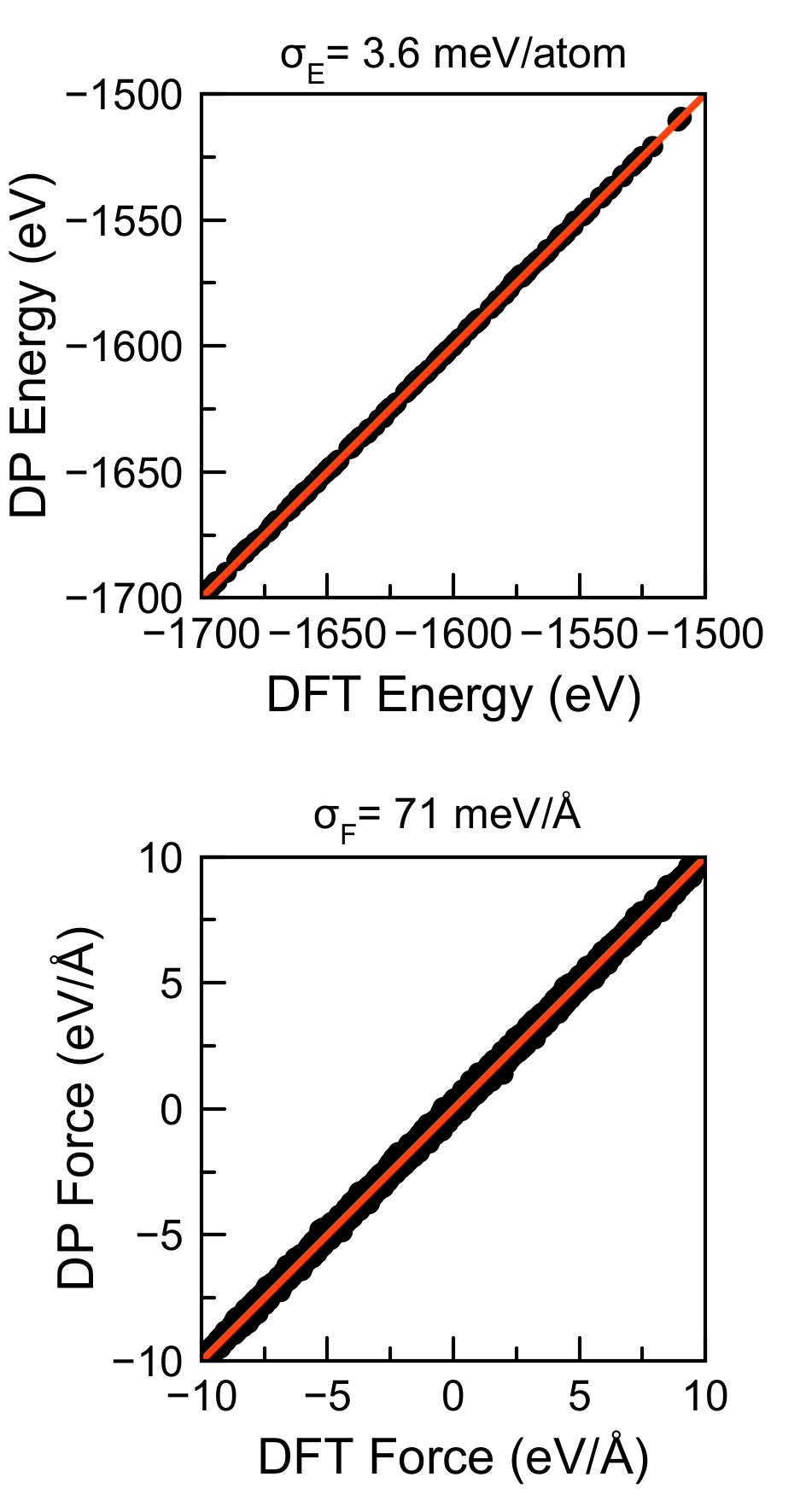}
	\caption{Comparison of potential energies and atomic forces from DP and DFT calculations.}
	\label{capvfigs1}
\end{figure}

After these DP-GEN iterations, the final training dataset comprises more than 3,000 configurations annotated with \textit{ab initio} force and energy information. Our DP-SCAN reaches an accuracy of $\sim$3.6 $\text{meV}/\text{atom}$ root mean squared error (RMSE) in energy and $\sim$71 \text{eV}/\text{\AA} RMSE in force compared to DFT-SCAN (Fig. \ref{capvfigs1}).

\newpage
\section{Supplementary Text S2: Ferroelasticity modelling}
\label{capvtext2}
Fig. \ref{capvfigs7} illustrates the applied stress ($\sigma_c$) and lattice parameters (\textbf{a},\textbf{b},\textbf{c}) during molecular dynamics (MD) simulations at 150 K and 12 GPa, highlighting a structural transition between a lying-down form (LF) and standing form (SF), as indicated by the background colors. The upper panel shows the normal stress along the \textbf{c}-axis ($\sigma_c$), which corresponds to the long side of the tetragonal phase, measured in GPa and plotted as a function of MD simulation time (ps). Initially, an external compressive stress was applied to the system, increasing gradually from 0 GPa to 5 GPa. After reaching this peak, the stress decreases until it reaches zero and further decreases, becoming tensile until a maximum tensile stress of 5 GPa is achieved. Finally, the applied tensile stress is gradually reduced until the system returns to its initial stress-free state. The switching between the two orientation states of CaPv, induced by the external stress $\sigma_c$, occurs at specific values of compressive and tensile stress. The lower panel shows the time-dependent lattice parameters (\textbf{a},\textbf{b},\textbf{c}) in angstroms. Notably, upon entering the lying-down form, the \textbf{a} and \textbf{b} parameters show significant divergence, with $\textbf{b} \neq \textbf{c}$. This occurs because the system transitions from a state where the \textbf{c}-axis is the long axis of the tetragonal phase ($\textbf{a} = \textbf{b} < \textbf{c}$) to one where the \textbf{a}-axis becomes the long axis ($ \textbf{b} = \textbf{c} < \textbf{a}$), caused by the applied compressive stress along the \textbf{c}-axis. The purple dashed line in the figure indicates that when $\sigma_c$ returns to zero, the system reverts to the $ \textbf{b} = \textbf{c} < \textbf{a}$ tetragonal phase. This suggests that the transition is characterized by a switch in the long axis of the tetragonal phase between the \textbf{a}- and \textbf{c}-axes.

\newpage

\begin{figure}
\centering
	\includegraphics[width=\textwidth]{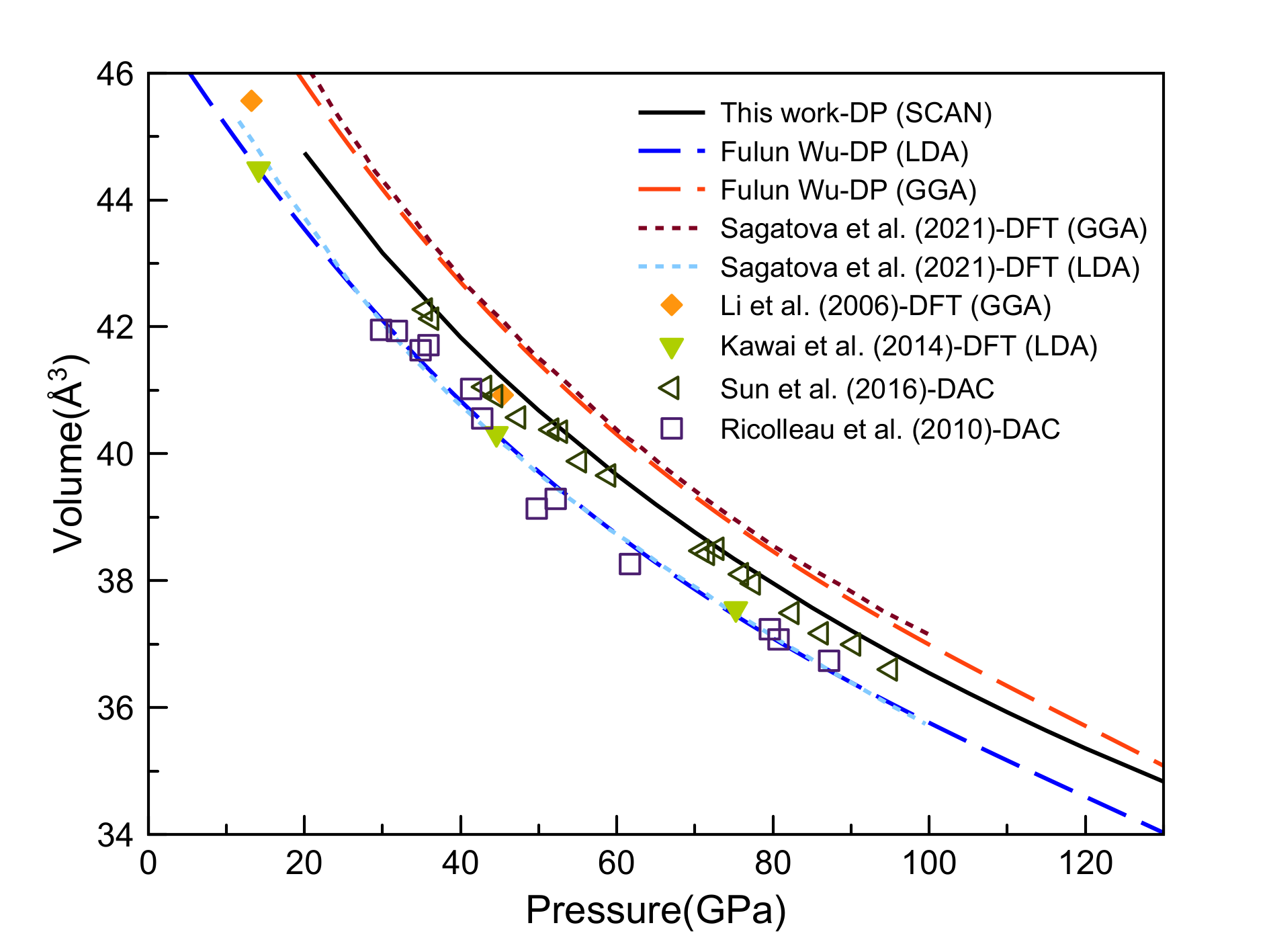}
	\caption{Compression curve of cubic CaPv at 2,000K. Solid curves are obtained using DP-SCAN in this work, and dashed lines are the previous calculations\cite{Kawai2014,Li2006b,Sagatova2021a,Wu2024a}. Previous measurements \cite{Ricolleau2010a,Sun2016} are also plotted for comparison.}
	\label{capvfigs2}
\end{figure}

\begin{figure}
\centering
	\includegraphics[width=\textwidth]{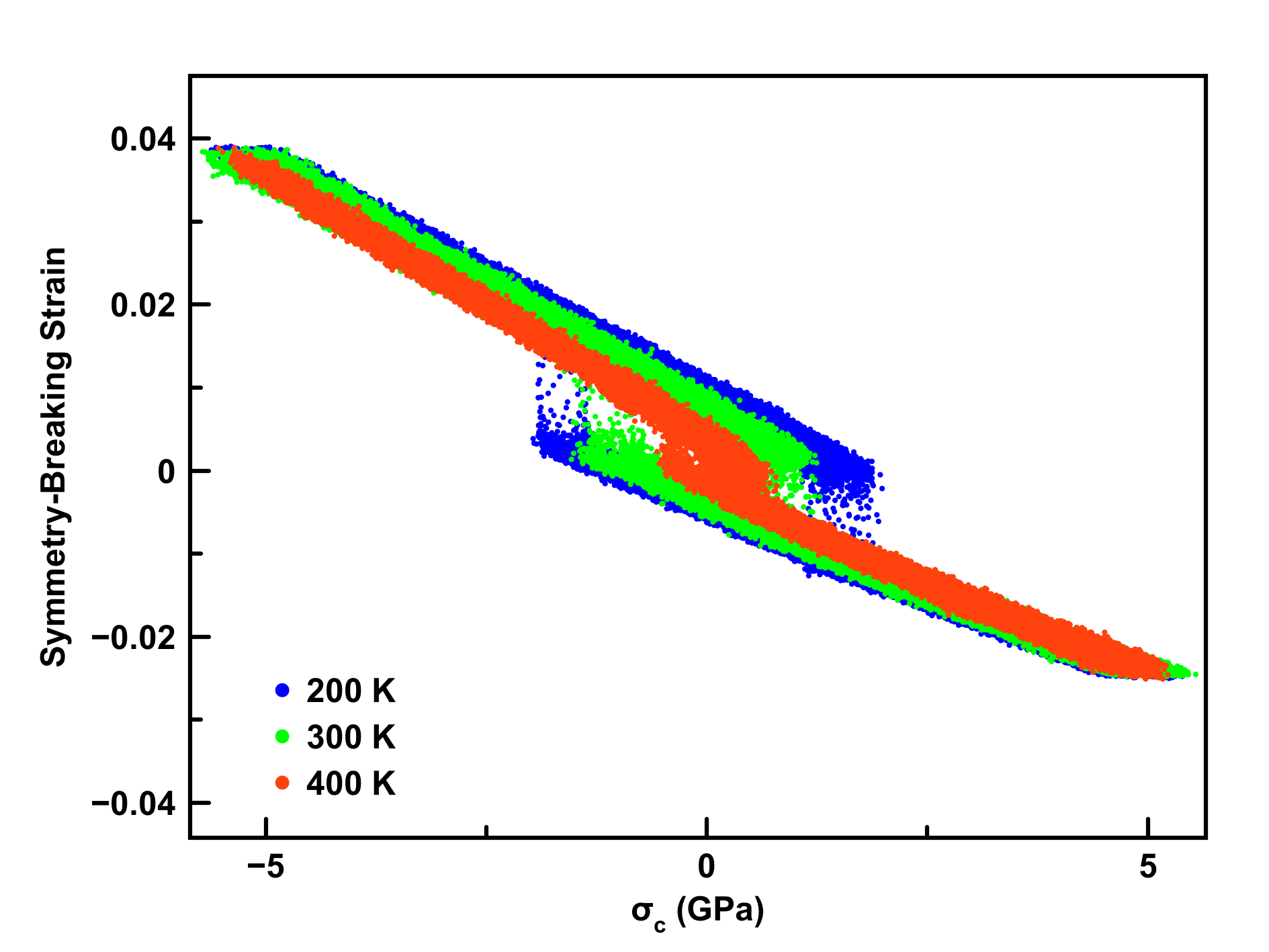}
	\caption{Temperature effect on the ferroelastic behavior of CaPv. The applied stress ($\sigma_c$) and symmetry-breaking strain $e_t=[(c-a)/a]$  during molecular dynamics (MD) simulations were obtained at 12 GPa and different temperatures. Symbol colors denote temperatures.}
	\label{capvfigs3}
\end{figure}

\begin{figure}
\centering
	\includegraphics[width=\textwidth]{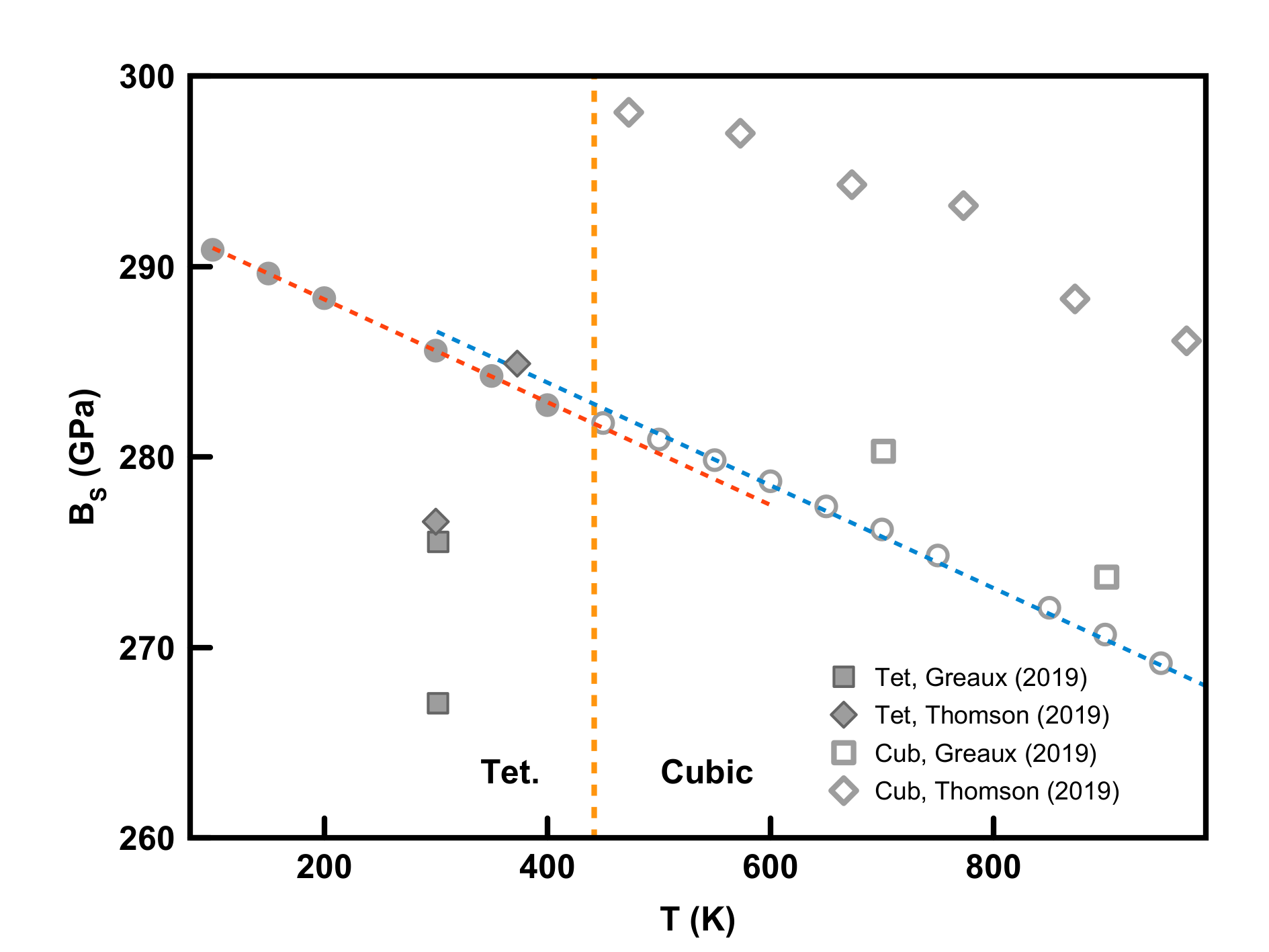}
	\caption{Bulk modulus-$B_S$ vs Temperature at 12 GPa. Filled symbols denote the tetragonal CaPv, while open symbols denote the cubic CaPv. Red and blue dashed lines are fitted linear curve for tetragonal and cubic CaPv, respectively. Measured elastic moduli from previous studies are plotted as grey diamonds \cite{Thomson2019a} (about 12 GPa) and grey squares \cite{Greaux2019a} (12 $\pm$ 1 GPa).Orange dashed lines indicates the phase transition line measured by Thomson et al. \cite{Thomson2019a}.}
	\label{capvfigs4}
\end{figure}

\begin{figure}
\centering
	\includegraphics[width=0.7\textwidth]{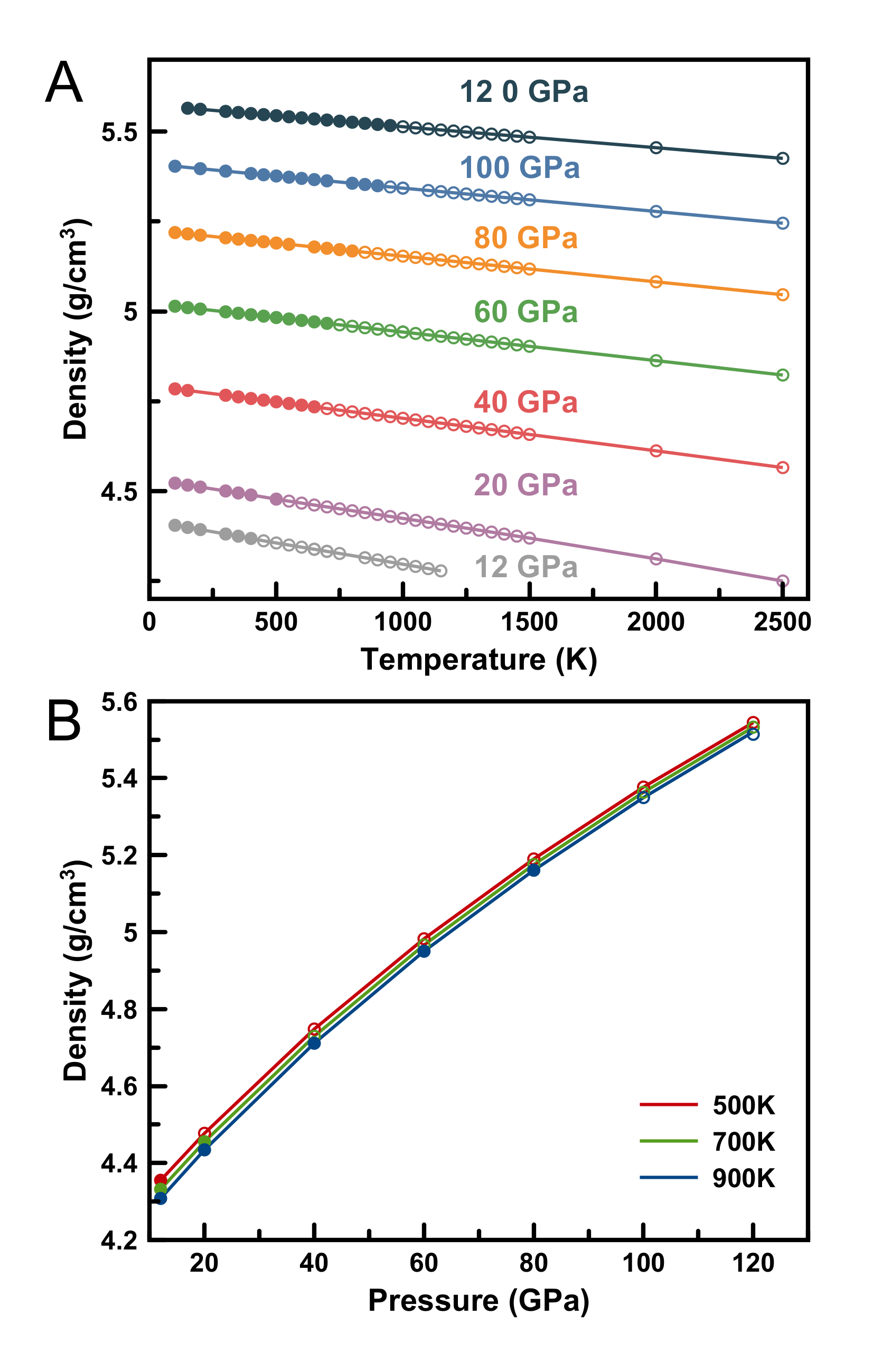}
	\caption{Density of CaPv across the ferroelastic\(\leftrightarrow \)paraelastic transition. (A) Density vs Temperature. Colors denote different pressures. (B) Density vs Pressure. Colors denote different temperatures. Filled symbols denote the tetragonal CaPv, while open symbols denote the cubic CaPv. }
	\label{capvfigs5}
\end{figure}

\begin{figure}
\centering
	\includegraphics[width=\textwidth]{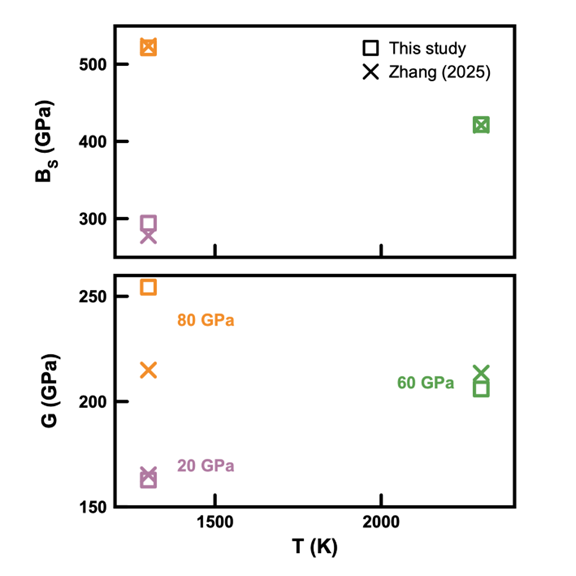}
	\caption{Comparison of bulk modulus and shear modulus vs temperature. Colors denote different pressures. Crosses denote the results from Zhang et al. \cite{Zhang2025}, while squares denote the results from this study. The difference at 80GPa results from the softening effect near the transition temperature, which is higher in Zhang et al. ’s result.}
	\label{capvfigs6}
\end{figure}

\begin{figure}
\centering
	\includegraphics[width=\textwidth]{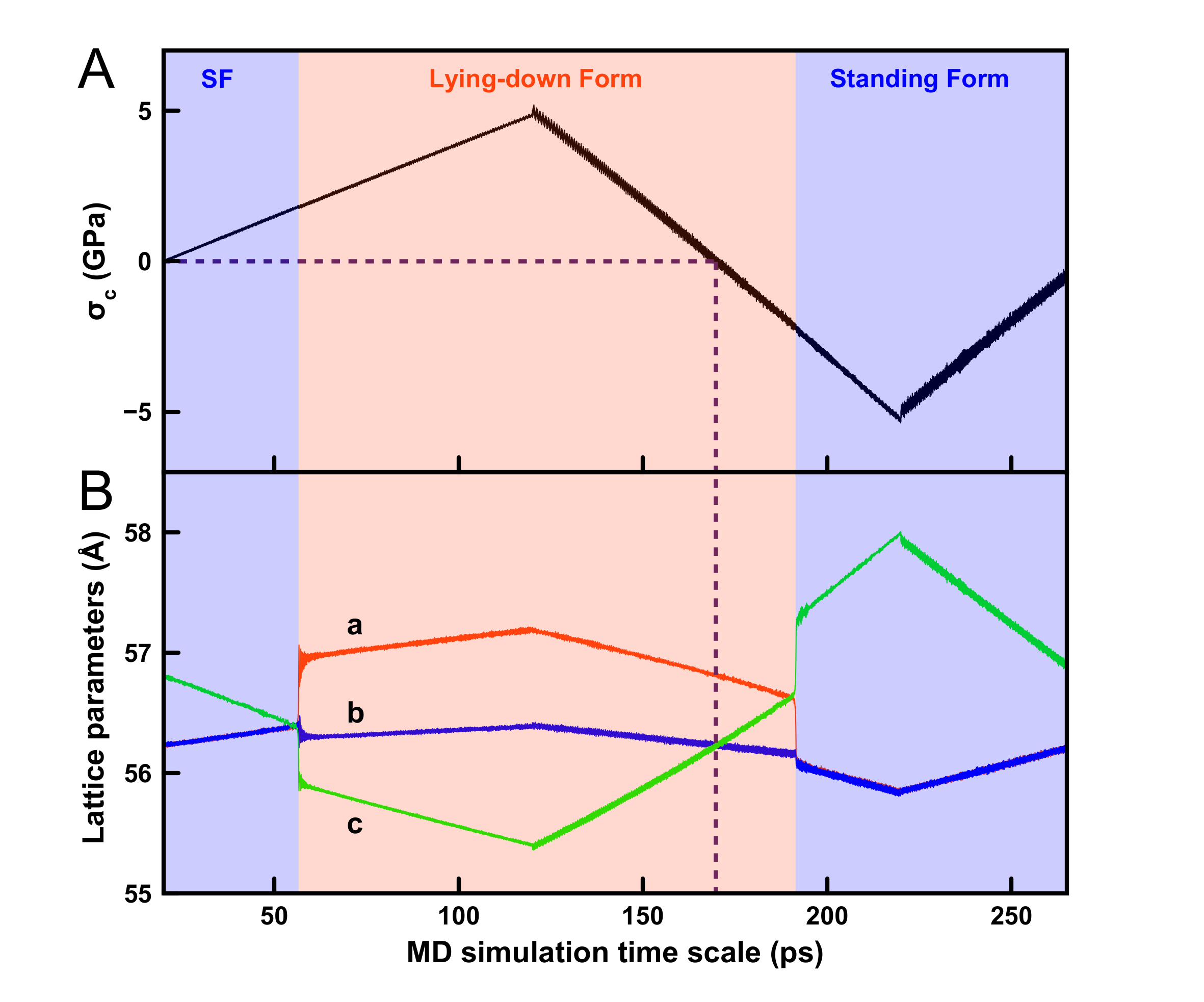}
	\caption{Evolution of stress ($\sigma_c$) and lattice parameters (\textbf{a},\textbf{b},\textbf{c}), where (\textbf{a},\textbf{b},\textbf{c}) represents the lattice vectors of the tetragonal primitive cell ($\sqrt{2} \times \sqrt{2} \times 2$, 20 atoms), during MD simulations at 12 GPa and 150 K. (A) The normal stress in GPa along the \textbf{c}-axis ($\sigma_c$) of the SF state was plotted as a function of MD simulation time. Purple dashed line indicates the stress-free state ($\sigma_c$=0); (B) The external stress ($\sigma_c$) dependence of the lattice parameters (\textbf{a},\textbf{b},\textbf{c}) in \text{\AA}.}
	\label{capvfigs7}
\end{figure}

\clearpage

\begin{table}[htbp]
\vspace{-3pt}
\caption{Thermoelastic properties of tetragonal CaPv and cubic CaPv}
\begin{tabular}{llllllll}
\Xhline{1px}
\rule{0pt}{4ex}
\raisebox{0.6ex}[0pt][0pt]{Ref.}                     & \raisebox{0.6ex}[0pt][0pt]{Phase} & \raisebox{0.6ex}[0pt][0pt]{$B_{S0}$ (GPa)} & \raisebox{0.6ex}[0pt][0pt]{$\frac{\partial B_S}{\partial P}$ }   & \raisebox{0.6ex}[0pt][0pt]{$\frac{\partial B_S}{\partial T}$(GPa/K)} & \raisebox{0.6ex}[0pt][0pt]{$G_0$(GPa)} & \raisebox{0.6ex}[0pt][0pt]{$\frac{\partial G}{\partial P}$}     & \raisebox{0.6ex}[0pt][0pt]{$\frac{\partial G}{\partial T}$(GPa/K)} \\
\hline
Theoretical calculations &       &       &     &         &       &      &         \\
This study               & Tet.  & 246.4 & 3.8 & -0.018  & 165.2 & 1.09 & -0.044  \\
                         & Cub.  & 253.6 & 3.8 & -0.018  & 168.2 & 1.55 & -0.016  \\
Ref. \cite{Tsuchiya2011}*                    & Tet.  & 247.3 & 3.7 & -       & 144.1 & 1.1  & -       \\
                         & Cub.  & 249.1 & 3.7 &         & 174.3 & 1.8  & -       \\
Experiments              &       &       &     &         &       &      &         \\
Ref. \cite{Greaux2019a}                   & Tet.  & 228   & 3.7 & -       & 116   & 1.3  & -       \\
                         & Cub.  & 248   & 4.2 & -0.036  & 126   & 1.6  & -0.015  \\
Ref. \cite{Sinelnikov1998}                   & Tet.  & 212   & -   & -       & 112   & -    & -       \\
Ref. \cite{Kudo2012}                    & Tet.  & -     & -   & -       & 115.8 & 1.2  &    \\
\Xhline{1px}
\end{tabular}
\end{table}

\noindent $K_{S0}$ and G$_{0}$ here were estimated at 300K for tetragonal CaPv and 700K for cubic CaPv.

\noindent * No temperature effects were taken into account in Ref. \cite{Tsuchiya2011}.

\bibliographystyle{apsrev4-1}

\begin{thebibliography}{60}
\bibitem{Irifune1994} Tetsuo Irifune, \href{https://doi.org/10.1038/370131a0}{Nature \textbf{370}, 131--133 (1994)}.
\bibitem{Saikia2008} Ashima Saikia, Daniel J. Frost, David C. Rubie, \href{https://doi.org/10.1126/science.1152818}{Science \textbf{319}, 1515--1518 (2008)}.
\bibitem{Hirose2002} Kei Hirose, Yingwei Fei, \href{https://doi.org/10.1016/S0016-7037(02)00847-5}{Geochimica et Cosmochimica Acta \textbf{66}, 2099--2108 (2002)}.
\bibitem{Irifune1993} T. Irifune, A. E. Ringwood, \href{https://doi.org/10.1016/0012-821X(93)90120-X}{Earth and Planetary Science Letters \textbf{117}, 101--110 (1993)}.
\bibitem{Kesson1994} S. E. Kesson, J. D. Fitz Gerald, J. M.G. Shelley, \href{https://doi.org/10.1038/372767a0}{Nature \textbf{372}, 767--769 (1994)}.
\bibitem{Irifune2000} T. Irifune, M. Miyashita, T. Inoue, J. Ando, K. Funakoshi, W. Utsumi, \href{https://doi.org/10.1029/2000GL012105}{Geophysical Research Letters \textbf{27}, 3541--3544 (2000)}.
\bibitem{Jung2011} Daniel Y. Jung, Max W. Schmidt, \href{https://doi.org/10.1007/s00269-010-0405-0}{Physics and Chemistry of Minerals \textbf{38}, 311--319 (2011)}.
\bibitem{Muir2021} Joshua M.R. Muir, Andrew R. Thomson, Feiwu Zhang, \href{https://doi.org/10.1016/j.epsl.2021.116973}{Earth and Planetary Science Letters \textbf{566}, 116973 (2021)}.
\bibitem{Vitos2006} L. Vitos, B. Magyari-Köpe, R. Ahuja, J. Kollár, G. Grimvall, B. Johansson, \href{https://doi.org/10.1016/j.pepi.2006.02.004}{Physics of the Earth and Planetary Interiors \textbf{156}, 108--116 (2006)}.
\bibitem{Ko2022} Byeongkwan Ko, Eran Greenberg, Vitali Prakapenka, E. Ercan Alp, Wenli Bi, Yue Meng, Dongzhou Zhang, Sang Heon Shim, \href{https://doi.org/10.1038/s41586-022-05237-4}{Nature \textbf{611}, 88--92 (2022)}.
\bibitem{Chao2024} Keng-Hsien Chao, Meryem Berrada, Siheng Wang, Juliana Peckenpaugh, Dongzhou Zhang, Stella Chariton, Vitali Prakapenka, Bin Chen, \href{https://doi.org/10.2138/am-2023-9104}{American Mineralogist \textbf{109}, 1861--1870 (2024)}.
\bibitem{Thomson2016} Andrew R. Thomson, Michael J. Walter, Simon C. Kohn, Richard A. Brooker, \href{https://doi.org/10.1038/nature16174}{Nature \textbf{529}, 76--79 (2016)}.
\bibitem{Thomson2019a} A. R. Thomson, W. A. Crichton, J. P. Brodholt, I. G. Wood, N. C. Siersch, J. M.R. Muir, D. P. Dobson, S. A. Hunt, \href{https://doi.org/10.1038/s41586-019-1483-x}{Nature \textbf{572}, 643--647 (2019)}.
\bibitem{Komabayashi2007} Tetsuya Komabayashi, Kei Hirose, Nagayoshi Sata, Yasuo Ohishi, Leonid S. Dubrovinsky, \href{https://doi.org/10.1016/j.epsl.2007.06.015}{Earth and Planetary Science Letters \textbf{260}, 564--569 (2007)}.
\bibitem{Kurashina2004} Tsuyoshi Kurashina, Kei Hirose, Shigeaki Ono, Nagayoshi Sata, Yasuo Ohishi, \href{https://doi.org/10.1016/j.pepi.2004.02.005}{Physics of the Earth and Planetary Interiors \textbf{145}, 67--74 (2004)}.
\bibitem{Caracas2005} R. Caracas, R. Wentzcovitch, G. David Price, J. Brodholt, \href{https://doi.org/10.1029/2004GL022144}{Geophysical Research Letters \textbf{32}, 1--5 (2005)}.
\bibitem{Li2006b} Li Li, Donald J. Weidner, John Brodholt, Dario Alfè, G. David Price, Razvan Caracas, Renata Wentzcovitch, \href{https://doi.org/10.1016/j.pepi.2005.12.007}{Physics of the Earth and Planetary Interiors \textbf{155}, 260--268 (2006)}.
\bibitem{Sagatova2021a} D. N. Sagatova, A. F. Shatskiy, N. E. Sagatov, K. D. Litasov, \href{https://doi.org/10.1134/S0016702921080073}{Geochemistry International \textbf{59}, 791--800 (2021)}.
\bibitem{Wu2024a} Fulun Wu, Yang Sun, Tianqi Wan, Shunqing Wu, Renata M. Wentzcovitch, \href{https://doi.org/10.1029/2023GL108012}{Geophysical Research Letters \textbf{51}, e2023GL108012 (2024)}.
\bibitem{Carpenter1998b} Michael A. Carpenter, Ekhard K.H. Salje, \href{https://doi.org/10.1127/ejm/10/4/0693}{European Journal of Mineralogy \textbf{10}, 693--812 (1998)}.
\bibitem{Salje1976} E. Salje, G. Hoppmann, \href{https://doi.org/10.1016/0025-5408(76)90107-0}{Materials Research Bulletin \textbf{11}, 1545--1549 (1976)}.
\bibitem{Zhang2021a} Yanyao Zhang, Suyu Fu, Baoyun Wang, Jung Fu Lin, \href{https://doi.org/10.1103/PhysRevLett.126.025701}{Physical Review Letters \textbf{126}, 025701 (2021)}.
\bibitem{Liu1975} Lin Gun Liu, A. E. Ringwood, \href{https://doi.org/10.1016/0012-821X(75)90229-0}{Earth and Planetary Science Letters \textbf{28}, 209--211 (1975)}.
\bibitem{Sun2015} Jianwei Sun, Adrienn Ruzsinszky, Johnp Perdew, \href{https://doi.org/10.1103/PhysRevLett.115.036402}{Physical Review Letters \textbf{115}, 36402 (2015)}.
\bibitem{Luo2024} Chenxing Luo, Yang Sun, Renata Wentzcovitch, \href{https://doi.org/10.1103/PhysRevMaterials.8.103601}{Physical Review Materials \textbf{8}, 103601 (2024)}.
\bibitem{Chen2018} Huawei Chen, Sang Heon Shim, Kurt Leinenweber, Vitali Prakapenka, Yue Meng, Clemens Prescher, \href{https://doi.org/10.2138/am-2018-6087}{American Mineralogist \textbf{103}, 462--468 (2018)}.
\bibitem{Greaux2019a} Steeve Gréaux, Tetsuo Irifune, Yuji Higo, Yoshinori Tange, Takeshi Arimoto, Zhaodong Liu, Akihiro Yamada, \href{https://doi.org/10.1038/s41586-018-0816-5}{Nature \textbf{565}, 218--221 (2019)}.
\bibitem{Mao1989} H. K. Mao, L. C. Chen, R. J. Hemley, A. P. Jephcoat, Y. Wu, W. A. Bassett, \href{https://doi.org/10.1029/jb094ib12p17889}{Journal of Geophysical Research \textbf{94}, 17889--17894 (1989)}.
\bibitem{Shim2002} Sang Heon Shim, Raymond Jeanloz, Thomas S. Duffy, \href{https://doi.org/10.1029/2002gl016148}{Geophysical Research Letters \textbf{29}, 2166 (2002)}.
\bibitem{Wang1996} Yanbin Wang, Donald J. Weidner, François Guyot, \href{https://doi.org/10.1029/95jb03254}{Journal of Geophysical Research: Solid Earth \textbf{101}, 661--672 (1996)}.
\bibitem{Ricolleau2010a} Angle Ricolleau, Jean Philippe Perrillat, Guillaume Fiquet, Isabelle Daniel, Jan Matas, Ahmed Addad, Nicolas Menguy, Herv Cardon, Mohamed Mezouar, Nicolas Guignot, \href{https://doi.org/10.1029/2009JB006709}{Journal of Geophysical Research: Solid Earth \textbf{115}, B08202 (2010)}.
\bibitem{Salje2012} Ekhard K.H. Salje, \href{https://doi.org/10.1146/annurev-matsci-070511-155022}{Annual Review of Materials Research \textbf{42}, 265--283 (2012)}.
\bibitem{Kaneshima2016} Satoshi Kaneshima, \href{https://doi.org/10.1016/j.pepi.2016.05.004}{Physics of the Earth and Planetary Interiors \textbf{257}, 105--114 (2016)}.
\bibitem{Salje1992} E. K.H. Salje, \href{https://doi.org/10.1016/0370-1573(92)90035-X}{Physics Reports \textbf{215}, 49--99 (1992)}.
\bibitem{Zhuang2024a} Jingyi Zhuang, Renata Wentzcovitch, \href{https://doi.org/10.1029/2024GL108967}{Geophysical Research Letters \textbf{51}, 1--18 (2024)}.
\bibitem{Brown1981} J. M. Brown, T. J. Shankland, \href{https://doi.org/10.1111/j.1365-246X.1981.tb04891.x}{Geophysical Journal of the Royal Astronomical Society \textbf{66}, 579--596 (1981)}.
\bibitem{Eberle2002} Michael A. Eberle, Olivier Grasset, Christophe Sotin, \href{https://doi.org/10.1016/S0031-9201(02)00157-7}{Physics of the Earth and Planetary Interiors \textbf{134}, 191--202 (2002)}.
\bibitem{Zhang2025} Chi Zhang , Jin-Yuan Yang , Tao Sun , Huai Zhang , John P. Brodholt , \href{https://doi.org/10.1073/pnas.2410910122}{Proceedings of the National Academy of Sciences \textbf{122}, e2410910122 (2025)}.
\bibitem{shin2024} Yongjoong Shin, Enrico Di Lucente, Nicola Marzari, Lorenzo Monacelli, arXiv \textbf{}, 2411.18489 (2025).
\bibitem{Stixrude2007} Lars Stixrude, C. Lithgow-Bertelloni, B. Kiefer, P. Fumagalli, \href{https://doi.org/10.1103/PhysRevB.75.024108}{Physical Review B - Condensed Matter and Materials Physics \textbf{75}, 024108 (2007)}.
\bibitem{Tsuchiya2011} Taku Tsuchiya, \href{https://doi.org/10.1016/j.pepi.2011.06.018}{Physics of the Earth and Planetary Interiors \textbf{188}, 142--149 (2011)}.
\bibitem{Dziewonski1981} Adam M. Dziewonski, Don L. Anderson, \href{https://doi.org/10.1016/0031-9201(81)90046-7}{Physics of the Earth and Planetary Interiors \textbf{25}, 297--356 (1981)}.
\bibitem{Wan2024} Tianqi Wan, Chenxing Luo, Yang Sun, Renata M. Wentzcovitch, \href{https://doi.org/10.1103/PhysRevB.109.094101}{Physical Review B \textbf{109}, 94101 (2024)}.
\bibitem{Brandenburg2007} J. P. Brandenburg, P. E. van Keken, \href{https://doi.org/10.1029/2006JB004813}{Journal of Geophysical Research: Solid Earth \textbf{112}, B06403 (2007)}.
\bibitem{McNamara2005} Allen K. McNamara, Shijie Zhong, \href{https://doi.org/10.1038/nature04066}{Nature \textbf{437}, 1136--1139 (2005)}.
\bibitem{Kohutych2010} A. Kohutych, R. Yevych, S. Perechinskii, V. Samulionis, J. Banys, Yu Vysochanskii, \href{https://doi.org/10.1103/PhysRevB.82.054101}{Physical Review B - Condensed Matter and Materials Physics \textbf{82}, 054101 (2010)}.
\bibitem{Carpenter2007} Michael A. Carpenter, \href{https://doi.org/10.2138/am.2007.2295}{American Mineralogist \textbf{92}, 309--327 (2007)}.
\bibitem{chenxing30} Kresse, G., Furthm\"uller, J., \href{https://doi.org/10.1103/PhysRevB.54.11169}{Phys. Rev. B \textbf{54}, 11169--11186 (1996)}.
\bibitem{Zhang2021c} Zhen Zhang, Renata M. Wentzcovitch, \href{https://doi.org/10.1103/PhysRevB.103.104108}{Physical Review B \textbf{103}, 104108 (2021)}.
\bibitem{Zhang2018b} Zhang, Linfeng, Han, Jiequn, Wang, Han, Saidi, Wissam A., Car, Roberto, Weinan, E., \href{https://proceedings.neurips.cc/paper_files/paper/2018/file/e2ad76f2326fbc6b56a45a56c59fafdb-Paper.pdf} {Adv. Neural Inf. Process. Syst.}  (2018).
\bibitem{DPMD} Zhang, Linfeng, Han, Jiequn, Wang, Han, Car, Roberto, E, Weinan, \href{https://doi.org/10.1103/PhysRevLett.120.143001}{Phys. Rev. Lett. \textbf{120}, 143001 (2018)}.
\bibitem{Thompson2022} Aidan P. Thompson, H. Metin Aktulga, Richard Berger, Dan S. Bolintineanu, W. Michael Brown, Paul S. Crozier, Pieter J. in 't Veld, Axel Kohlmeyer, Stan G. Moore, Trung Dac Nguyen, Ray Shan, Mark J. Stevens, Julien Tranchida, Christian Trott, Steven J. Plimpton, \href{https://doi.org/10.1016/j.cpc.2021.108171}{Computer Physics Communications \textbf{271}, 108171 (2022)}.
\bibitem{Hoover1996} William G. Hoover, Brad Lee Holian, \href{https://doi.org/10.1016/0375-9601(95)00973-6}{Physics Letters, Section A: General, Atomic and Solid State Physics \textbf{211}, 253--257 (1996)}.
\bibitem{Wentzcovitch1991a} Renata M. Wentzcovitch, \href{https://doi.org/10.1103/PhysRevB.44.2358}{Physical Review B \textbf{44}, 2358--2361 (1991)}.
\bibitem{Clavier2023} Germain Clavier, Aidan P. Thompson, \href{https://doi.org/10.1016/j.cpc.2023.108674}{Computer Physics Communications \textbf{286}, 108674 (2023)}.
\bibitem{Ray1984} John R. Ray, Aneesur Rahman, \href{https://doi.org/10.1063/1.447221}{The Journal of Chemical Physics \textbf{80}, 4423--4428 (1984)}.
\bibitem{Zhen2012} Yubao Zhen, Chengbiao Chu, \href{https://doi.org/10.1016/j.cpc.2011.09.006}{Computer Physics Communications \textbf{183}, 261--265 (2012)}.
\bibitem{Barron1965} T. H.K. Barron, M. L. Klein, \href{https://doi.org/10.1088/0370-1328/85/3/313}{Proceedings of the Physical Society \textbf{85}, 523--532 (1965)}.
\bibitem{Luo2022} Chenxing Luo, Jeroen Tromp, Renata M. Wentzcovitch, \href{https://doi.org/10.1103/PhysRevB.106.214104}{Physical Review B \textbf{106}, 214104 (2022)}.





\end{thebibliography}

\begin{thebibliography}{99}
\bibitem{Zhang2018b} Zhang, Linfeng, Han, Jiequn, Wang, Han, Saidi, Wissam A., Car, Roberto, Weinan, E., \href{https://proceedings.neurips.cc/paper_files/paper/2018/file/e2ad76f2326fbc6b56a45a56c59fafdb-Paper.pdf} {Adv. Neural Inf. Process. Syst.}  (2018).
\bibitem{Wang2018} Han Wang, Linfeng Zhang, Jiequn Han, Weinan E, \href{https://doi.org/10.1016/j.cpc.2018.03.016}{Computer Physics Communications \textbf{228}, 178--184 (2018)}.
\bibitem{chenxing37} Zeng, Jinzhe, Zhang, Duo, Lu, Denghui, Mo, Pinghui, Li, Zeyu, Chen, Yixiao, Rynik, Marián, Huang, Li’ang, Li, Ziyao, Shi, Shaochen, Wang, Yingze, Ye, Haotian, Tuo, Ping, Yang, Jiabin, Ding, Ye, Li, Yifan, Tisi, Davide, Zeng, Qiyu, Bao, Han, Xia, Yu, Huang, Jiameng, Muraoka, Koki, Wang, Yibo, Chang, Junhan, Yuan, Fengbo, Bore, Sigbjørn Løland, Cai, Chun, Lin, Yinnian, Wang, Bo, Xu, Jiayan, Zhu, Jia-Xin, Luo, Chenxing, Zhang, Yuzhi, Goodall, Rhys E. A., Liang, Wenshuo, Singh, Anurag Kumar, Yao, Sikai, Zhang, Jingchao, Wentzcovitch, Renata, Han, Jiequn, Liu, Jie, Jia, Weile, York, Darrin M., E, Weinan, Car, Roberto, Zhang, Linfeng, Wang, Han, \href{https://doi.org/10.1063/5.0155600}{The Journal of Chemical Physics \textbf{159}, 054801 (2023)}.
\bibitem{Kingma2014} Diederik P. Kingma, Jimmy Lei Ba, 3rd International Conference on Learning Representations, ICLR 2015 - Conference Track Proceedings \textbf{},  (2015).
\bibitem{Zhang2020} Yuzhi Zhang, Haidi Wang, Weijie Chen, Jinzhe Zeng, Linfeng Zhang, Han Wang, Weinan E, \href{https://doi.org/10.1016/j.cpc.2020.107206}{Computer Physics Communications \textbf{253}, 107206 (2020)}.
\bibitem{Zhang2019} Linfeng Zhang, De Ye Lin, Han Wang, Roberto Car, E. Weinan, \href{https://doi.org/10.1103/PhysRevMaterials.3.023804}{Physical Review Materials \textbf{3}, 23804 (2019)}.
\bibitem{Wu2024a} Fulun Wu, Yang Sun, Tianqi Wan, Shunqing Wu, Renata M. Wentzcovitch, \href{https://doi.org/10.1029/2023GL108012}{Geophysical Research Letters \textbf{51}, e2023GL108012 (2024)}.
\bibitem{Kawai2014} Kenji Kawai, Taku Tsuchiya, \href{https://doi.org/10.1002/2013JB010905}{Journal of Geophysical Research: Solid Earth \textbf{119}, 2801--2809 (2014)}.
\bibitem{Li2006b} Li Li, Donald J. Weidner, John Brodholt, Dario Alfè, G. David Price, Razvan Caracas, Renata Wentzcovitch, \href{https://doi.org/10.1016/j.pepi.2005.12.007}{Physics of the Earth and Planetary Interiors \textbf{155}, 260--268 (2006)}.
\bibitem{Sagatova2021a} D. N. Sagatova, A. F. Shatskiy, N. E. Sagatov, K. D. Litasov, \href{https://doi.org/10.1134/S0016702921080073}{Geochemistry International \textbf{59}, 791--800 (2021)}.
\bibitem{Ricolleau2010a} Angle Ricolleau, Jean Philippe Perrillat, Guillaume Fiquet, Isabelle Daniel, Jan Matas, Ahmed Addad, Nicolas Menguy, Herv Cardon, Mohamed Mezouar, Nicolas Guignot, \href{https://doi.org/10.1029/2009JB006709}{Journal of Geophysical Research: Solid Earth \textbf{115}, B08202 (2010)}.
\bibitem{Sun2016} Ningyu Sun, Zhu Mao, Shuai Yan, Xiang Wu, Vitali B. Prakapenka, Jung Fu Lin, \href{https://doi.org/10.1002/2016JB013062}{Journal of Geophysical Research: Solid Earth \textbf{121}, 4876--4894 (2016)}.
\bibitem{Thomson2019a} A. R. Thomson, W. A. Crichton, J. P. Brodholt, I. G. Wood, N. C. Siersch, J. M.R. Muir, D. P. Dobson, S. A. Hunt, \href{https://doi.org/10.1038/s41586-019-1483-x}{Nature \textbf{572}, 643--647 (2019)}.
\bibitem{Greaux2019a} Steeve Gréaux, Tetsuo Irifune, Yuji Higo, Yoshinori Tange, Takeshi Arimoto, Zhaodong Liu, Akihiro Yamada, \href{https://doi.org/10.1038/s41586-018-0816-5}{Nature \textbf{565}, 218--221 (2019)}.
\bibitem{Zhang2025} Chi Zhang , Jin-Yuan Yang , Tao Sun , Huai Zhang , John P. Brodholt , \href{https://doi.org/10.1073/pnas.2410910122}{Proceedings of the National Academy of Sciences \textbf{122}, e2410910122 (2025)}.
\bibitem{Tsuchiya2011} Taku Tsuchiya, \href{https://doi.org/10.1016/j.pepi.2011.06.018}{Physics of the Earth and Planetary Interiors \textbf{188}, 142--149 (2011)}.
\bibitem{Sinelnikov1998} Y. D. Sinelnikov, G. Chen, R. C. Liebermann, \href{https://doi.org/10.1007/S002690050143/METRICS}{Physics and Chemistry of Minerals \textbf{25}, 515--521 (1998)}.
\bibitem{Kudo2012} Yuki Kudo, Kei Hirose, Motohiko Murakami, Yuki Asahara, Haruka Ozawa, Yasuo Ohishi, Naohisa Hirao, \href{https://doi.org/10.1016/j.epsl.2012.06.040}{Earth and Planetary Science Letters \textbf{349-350}, 1--7 (2012)}.

\end{thebibliography}

\end{document}